\documentclass[12pt,preprint]{aastex}

\usepackage{graphicx}
\usepackage{epstopdf}
\newcommand{\HI}{H$\,${\sc i}}

\begin{document}

\title{Feedback and galaxy dynamics: A study of turbulence and star formation in 34 galaxies using the PHANGS survey}

\author{Bruce G.\ Elmegreen\altaffilmark{1}}

\altaffiltext{1}{Katonah, New York USA, belmegreen@gmail.com}

\begin{abstract}
The correlation between interstellar turbulent speed and local star formation rate surface density, $\Sigma_{\rm SFR}$, is studied using CO observations in the PHANGS survey. The local velocity dispersion of molecular gas, $\sigma$, increases with $\Sigma_{\rm SFR}$, but the virial parameter, $\alpha_{\rm vir}$, is about constant, suggesting the molecular gas remains self-gravitating.  The correlation arises because $\sigma$ depends on the molecular surface density, $\Sigma_{\rm mol}$, and object cloud mass, $M_{\rm mol}$, with the usual molecular cloud correlations, while $\Sigma_{\rm SFR}$ increases with both of these quantities because of a nearly constant star formation efficiency for CO.  Pressure fluctuations with $\Delta \Sigma_{\rm SFR}$ are also examined. Azimuthal variations of molecular pressure, $\Delta P_{\rm mol}$, have a weaker correlation with $\Delta\Sigma_{\rm SFR}$  than expected from the power-law correlation between the total quantities, suggesting slightly enhanced SFR efficiency per molecule in spiral arms. Dynamical equilibrium pressure and star formation rate correlate well for the whole sample, as $P_{\rm DE}\propto\Sigma_{\rm SFR}^{1.3}$, which is steeper than in other studies. The azimuthal fluctuations, $\Delta P_{\rm DE}(\Delta\Sigma_{\rm SFR})$, follow the total correlation $P_{\rm DE}(\Sigma_{\rm SFR})$ closely, hinting that some of this correlation may be a precursor to star formation, rather than a reaction. Galactic dynamical processes correlate linearly such that $\Sigma_{\rm SFR}\propto(\Sigma_{\rm gas}R)^{1.0\pm0.3}$ for total gas surface density $\Sigma_{\rm gas}$ and galactic dynamical rates, $R$, equal to $\kappa$, $A$, or $\Omega$, representing epicyclic frequency, shear rate $A$, and orbit rate $\Omega$.  These results suggest important roles for both feedback and galactic dynamics.
\end{abstract}

\section{Introduction} \label{sec-intro}

Interstellar turbulence is driven by a combination of pressure from young stars, gravitational and magnetic instabilities, and interactions with galactic-scale processes such as spiral arms, bars, accretion, and neighboring galaxies. The proportion of each depends on their relative energy and momentum densities, which vary from galaxy to galaxy and within each galaxy \citep[see reviews in][]{elmegreen04,maclow04,kritsuk17,hayward17,burkhart21}. 

Star formation is also driven by a combination of processes \citep[see reviews in][]{mckee07,krumholz14,padoan14}. The rate generally depends on the gas mass and density, with molecular cloud cores ultimately undergoing gravitational collapse \citep{jin17,vazquez19,rosen20}.  Additional stimuli come from some of the same processes that generate turbulence, like pressures from young stars \citep[e.g.,][]{egorov23}, turbulent flow convergence and cloud collisions \citep[e.g.,][]{hartmann01,suwannajak14,wu18,fukui21}, and spiral shocks \citep{kO06,elmegreen19}. There are also impediments to collapse such as ISM stabilization by rotation and turbulence \citep{martig08}, cloud heating and disruption by young stars \citep[e.g.,][]{haid19,chevance20,lu20,grudic22}, and density restructuring by shear and cloud collisions in turbulent and galactic-scale flows \citep[][and references therein]{federrath15,hayward17}.

All of these processes interact in complex ways, producing both positive and negative feedback between turbulence and star formation. For example, low turbulent speed leads to a more unstable disk, which pumps energy back into the turbulence directly \citep{fleck81,bertin97,huber02,wada02} via swing-amplifier instabilities \citep{toomre91} and the resulting torque-driven disk inflows \citep{krumholz18}, and also increases turbulence indirectly through the resulting enhanced star formation rate (SFR) and its accompanying pressures \citep{goldreich65}.  Computer simulations of these interactions generally agree with observations on a wide range of scales \citep[e.g.][]{dobbs14,hung19,ostriker22}.    

The combined processes of turbulence and star formation may be viewed as comprising two main themes: (1) feedback control of star formation through ISM pressurization, disk thickening, and stabilization against gravitational collapse, with the same feedback driving turbulence, and (2) gravity-controlled star formation through swing-amplified instabilities, spiral-arm torques, and in-plane accretion tapping galactic potential energy as a source of turbulence. Both consider molecular cloud internal structure and turbulence as a result of their formation history and evolution, including collapse and the smaller-scale effects of feedback. Attempts at comparing these two themes through their scaling relations are usually inconclusive  \citep[e.g.,][]{suwannajak14,johnson18}, probably because there is a mixture of them in most regions and other processes operate too. For example, triggered cloud and star formation in stellar spiral waves might be considered a third theme, but it is difficult to recognize in global studies because the net effect on the SFR is small \citep{elmegreen86,kim20} even though the effect on star formation morphology is large \citep{elmegreen18}.

\cite{ostriker10} proposed a ``pressure-regulated feedback model,'' which predicts that young stars  generate turbulence on the scale of the disk thickness with a velocity dispersion, $\sigma$, that is independent of the SFR density $\Sigma_{\rm SFR}$ \cite[see also][]{joung09,faucher13}. This follows from a balance between the dissipation and generation rates of vertical momentum when $\Sigma_{\rm SFR}$ scales with the gas freefall rate at the midplane. Also in this case, $\Sigma_{\rm SFR}$ closely follows the total pressure \citep{ostriker11,kko11,shetty12,kim13,ostriker22}.  This prediction seems to be at variance with observations inside or among local galaxies, where the velocity dispersion or kinetic energy density appear to increase with the SFR \citep{tamburro09,stilp13}, but after removing the average radial profiles of these quantities, \cite{hunter21} and \cite{elmegreen22} showed for local dwarf irregulars and spiral galaxies that the  prediction is verified: the \HI\ velocity dispersion and kinetic energy density in each region of star formation is independent of any excess or deficit in the SFR compared to the average at that radius. We return to this method here for the molecular ISM. 

Increasing $\sigma$ with $\Sigma_{\rm SFR}$ is found for starbursts and high redshift galaxies \citep{lehnert09,lehnert13,swinbank12,green14,johnson18,oliva18} (but not always, see Genzel et al. 2011), but the turbulent speeds are much larger than in local galaxies and could be the result of different processes, such as accretion energy \citep{elmegreen10}, feedback from accretion-driven star formation \citep{hung19}, and gravitational instabilities \citep{krumholz18}.

In contrast to the lack of a correlation between fluctuations in $\sigma$ and $\Sigma_{\rm SFR}$ for local galaxies, there is a correlation between $\sigma$ and the previous star formation rate. \cite{stilp13} found in 17 dwarf irregular galaxies that \HI\ velocity dispersion correlates with $\Sigma_{\rm SFR}$ from 30 to 40 Myr ago. Similarly, \cite{hunter22} studied 4 dwarfs and found a correlation between \HI\ dispersion on 400 pc scales and $\Sigma_{\rm SFR}$ 100 to 200 Myr ago. \cite{hunter23} also found a 70 to 140 Myr delay for the correlation in another dwarf and showed it occurred primarily on a scale of 400 pc. Galaxy simulations suggest the same type of delay \citep{orr20}. 

The second prediction of the pressure-regulated feedback model, that $\Sigma_{\rm SFR}$ scales with total pressure for dynamical equilibrium, $P_{\rm DE}$ (see Eq. \ref{eq:pde} below), has been confirmed by observations in several studies, where it is written as a power law relation between the two quantities, with a power $d\log{\Sigma_{\rm SFR}}/d\log{P_{\rm DE}}$ slightly larger than unity, e.g., 1.21 in \cite{ostriker22}. We examine this relationship again here, getting a similarly good correlation but with a slope less than unity, $\sim 0.77$. We also find this correlation  in the azimuthal fluctuations of $P_{\rm DE}$ with $\Sigma_{\rm SFR}$, and that the slope of these fluctuations equals the slope predicted from the power law correlation between the total quantities. This equivalence implies that the quantities mutually vary in azimuth according to the same power law as the total correlation and contain no obvious time delays on the resolution scale (Sect. \ref{total}).  We also point out that $P_{\rm DE}$ is the sum of two terms that are commonly used separately for the Kennicutt-Schmidt (KS) \citep{kennicutt12} and ``Extended'' KS \citep{shi18} relations, suggesting that $P_{\rm DE}(\Sigma_{\rm SFR})$ variations might also be a predecessor to star formation, like gas in the KS relations. 

The influence of galactic-scale processes on SFR has also been examined numerous times with mixed conclusions. With a galactic-scale rate $R$ for conversion of gas into stars, we might expect $\Sigma_{\rm SFR}\propto\Sigma_{\rm gas}R$ for total gas surface density $\Sigma_{\rm gas}$. \cite{leroy08} found a steeper result, $\Sigma_{\rm SFR}/\Sigma_{\rm gas}\sim R^2$ for $R=\Omega$, the orbital rate.  \cite{suwannajak14} determine fit parameters for 16 nearby galaxies considering rates $R$ equal to the orbital rate $\Omega$, and something like the shear rate, $\Omega(1-0.7\beta)$. (The Oort shear rate used below is $A=0.5\Omega(1-\beta)$; $\beta$ is the local power-law slope of the rotation curve, $d\log V_{\rm rot}/d\log R$, or $d\log\Omega/d\log R+1$). \cite{suwannajak14} do not give fitted slopes for these relationships, but assume the slope is unity and derive the rms deviations. \cite{sun23} plotted $\Sigma_{\rm SFR}$ versus $\Sigma_{\rm mol}R$ for molecular surface density $\Sigma_{\rm mol}$ instead of total gas surface density and for $R$ equal to the orbital rate and the cloud-scale free-fall rate. The slopes they found are shallower than unity because the rates get larger for the inner and denser regions of galaxies, where $\Sigma_{\rm SFR}$ is larger, and the $\Sigma_{\rm SFR}(\Sigma_{\rm mol})$ relation alone is about linear without the rate multiplier.  Here we plot $\Sigma_{\rm SFR}$ versus $\Sigma_{\rm gas}R$ for total gas because galactic dynamical processes act on the total gas, and for epicyclic, shear, and orbital rates, $R$, finding essentially unity slopes for all of them.  

The organization of this paper is as follows. Section \ref{fluctuations} considers turbulent motions in the molecular medium where stars form and feedback is strongest. We examine PHANGS data \citep{leroy21} for correlations between CO velocity dispersion $\sigma$, molecular surface density $\Sigma_{\rm mol}$, molecular cloud mass $M_{\rm mol}$, and local star formation rate, $\Sigma_{\rm SFR}$, including also the dimensionless virial parameter $\alpha_{\rm vir}$. We study both excursions around the average radial profiles, as done previously for \HI\ to measure local effects \citep{hunter21,elmegreen22}, and the total quantities (Sect. \ref{total}) to compare with the \cite{larson81} laws for molecular clouds and conventional star-formation efficiencies in molecular gas \citep{krumholz19}.  Because individual molecular clouds are seldom resolved in other galaxies, we consider both the pixel and object versions of the PHANGS data compiled by \cite{sun22}. The pixel version gives molecular quantities measured on 150 pc scales and averaged over 1.5 kpc hexagons covering a galaxy disk, and the object versions are from fits to individual clouds, also averaged over the 1.5 kpc hexagons. We make a distinction between galaxies with strong and weak spiral arms, as determined by the  range of azimuthal variations in star formation rate density. We include 17 galaxies of each type, all with stellar masses above $2\times10^{10}\;M_\odot$, to get enough data points in each disk. 

To examine the second aspect of the two models mentioned above, we consider variations in molecular pressure and total dynamical pressure with the local star formation rate, again relative to their azimuthal averages and in total (Sect. \ref{generalP}). Similarities between the $P_{\rm DE}(\Sigma_{\rm SFR})$ relation, which suggest that feedback influences $P_{\rm DE}$ because of star formation, and the inverse but equivalent relation $\Sigma_{\rm SFR}(P_{\rm DE})$, which suggests that star formation follows the gas and associated dynamical rates, are discussed in Section \ref{reflect}. Finally, we consider how $\Sigma_{\rm SFR}$ varies locally and in total with the three main gas dynamical rates for large-scale processes, namely, the epicyclic, shear and orbital rates (Sect. \ref{dynamics}). The conclusions are in Section \ref{conclusions}.

\section{Parameter variations with azimuth in the PHANGS data for large galaxies}
\label{fluctuations}

The PHANGS survey \citep{leroy21} includes galaxy-wide maps of \HI, $\Sigma_{\rm SFR}$ as determined from several methods, and CO(2-1) using several formulae for CO-to-gas conversion. Viral parameters, interstellar pressures and other secondary quantities are also tabulated. Here we use version ``v4p0\_public\_release'' in tabular form \footnote{https:/www.canfar.net/storage/list/phangs/RELEASES/Sun\_etal\_2022}, as explained in \cite{sun22}.  The $\alpha_{\rm CO}$ conversion used is for the fiducial case, which has a metallicity dependence. An $\alpha_{\rm CO}$ that depends inversely with velocity dispersion \citep{chiang23} might be better, but the difference should not be large \citep{teng23}. 

\begin{figure}
\includegraphics[width=12cm]{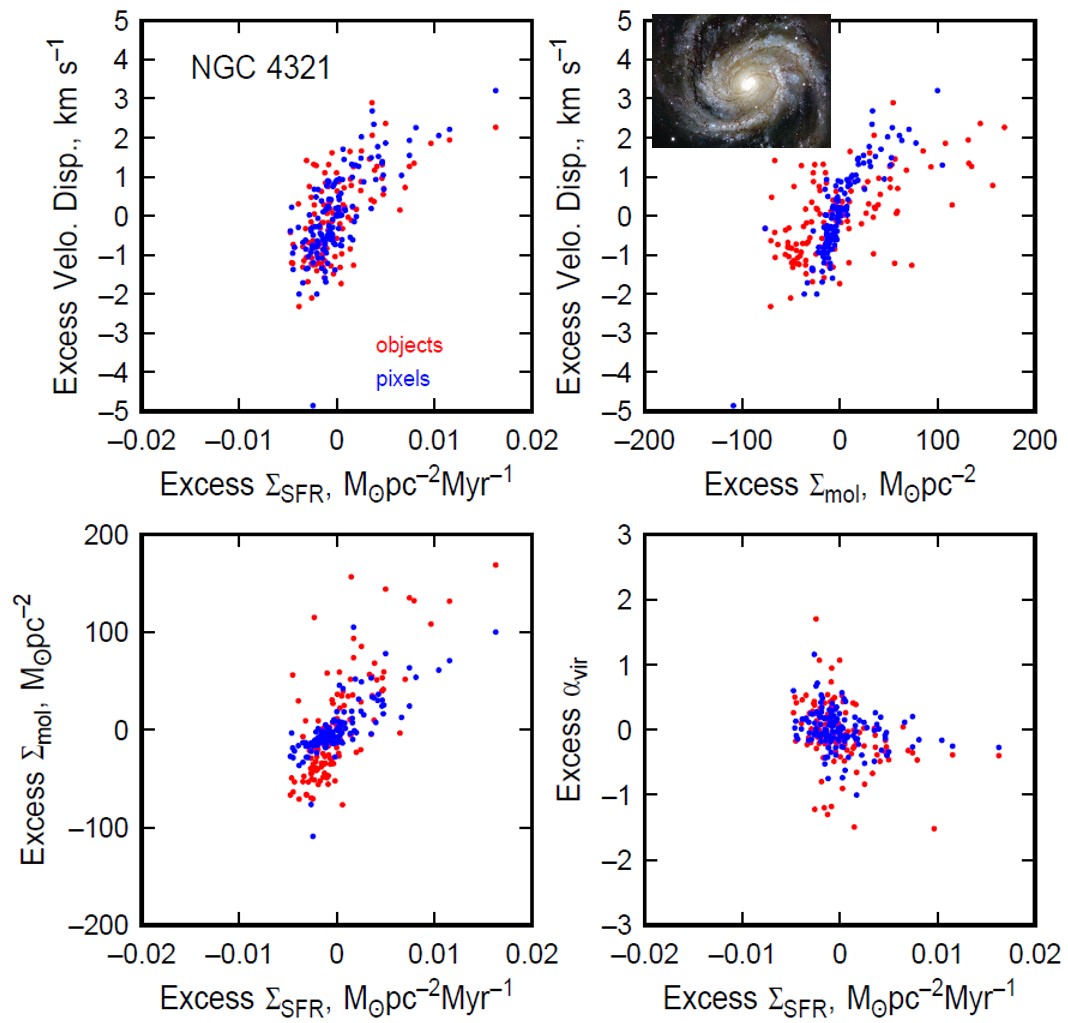}
\caption{Correlations between azimuthal fluctuations in various measured quantities for the strong-arm galaxy NGC 4321, from the PHANGS survey. ``Excess'' refers to the difference between the quantity and the average at that radius. The molecular surface density, $\Sigma_{\rm mol}$, and star formation surface density, $\Sigma_{\rm SFR}$, vary together above and below the azimuthal averages, giving a positive slope in the lower left panel. Excess velocity dispersion also varies with excess $\Sigma_{\rm SFR}$ (top left) and with excess $\Sigma_{\rm mol}$ (top right). The virial parameter, $\alpha_{\rm vir}$ does not change much with $\Sigma_{\rm SFR}$. Red points correspond to parameter measurements in 150 pc regions centered on individual molecular clouds, and blue points correspond to parameter measurements in a regular 150 pc pixel grid, both averaged within a 1.5 kpc hexagon, as tabulated by the PHANGS group.. An image of the galaxy is shown in the insert. }
\label{n4321-deltas}
\centering
\end{figure}

Figure \ref{n4321-deltas} shows correlated fluctuations between several quantities for the grand-design galaxy NGC  4321, shown in the insert. These are all fluctuations in azimuth, meaning that the average value of the quantity at that radius has been subtracted. In the various panels, the excess molecular surface density $\Sigma_{\rm mol}$ is shown versus the excess SFR density, $\Sigma_{\rm SFR}$ (lower left), the excess CO velocity dispersion $\sigma$ is shown versus the excess $\Sigma_{\rm SFR}$ (upper left) and $\Sigma_{\rm mol}$ (upper right), and the excess virial parameter $\alpha_{\rm vir}$ is shown versus the excess $\Sigma_{\rm SFR}$ (lower left). For the discussion, we denote excess quantities by the symbol $\Delta$ and we denote a correlation of A versus B by A(B). The figure shows a positive correlation for $\Delta\sigma(\Delta\Sigma_{\rm SFR})$ in the sense that regions with higher than the azimuthally-averaged $\Sigma_{\rm SFR}$ at their radius also have higher $\sigma$. We note there are not many regions with low $\Delta \Sigma_{\rm SFR}$ or low pixel-wise measurements of $\Delta\Sigma_{\rm mol}$ (blue dots in the top-right) for NGC 4321, although there are sub-zero $\Delta\sigma$ values. This is because star formation is highly concentrated in a small fraction of each annulus where there are spiral arms, so the increases above average are strong but atypical and the decreases below average are close to the average. The other galaxies studied here do not generally show such an asymmetry in $\Delta\Sigma_{\rm SFR}$.

The $\Delta \sigma(\Delta\Sigma_{\rm SFR}))$ correlation in the top-left panel might be a signature for turbulence enhancement by star formation feedback, but the other panels suggest that feedback contributions to $\sigma$ are minor: $\Delta\Sigma_{\rm mol}(\Delta\Sigma_{\rm SFR})$ (bottom left) is the usual molecular KS law and $\Delta\sigma(\Delta\Sigma_{\rm mol})$ (top right) is one of the usual Larson Laws for a fixed pixel size, so the $\Delta\sigma(\Delta\Sigma_{\rm SFR})$ correlation could be driven by pre-existing turbulence through the Larson Law, with regions of higher SFR following the presence of more molecular mass according to the KS law.  This interpretation will be quantified in several ways in Section \ref{total}.

The excess virial parameter in the lower right of Figure \ref{n4321-deltas} is not increasing as $\Delta \Sigma_{\rm SFR}$ increases, which might be expected if the molecular gas is dispersing because of feedback. For NGC 4321, $\Delta \alpha_{\rm vir}$ decreases a little with increasing $\Delta\Sigma_{\rm SFR}$.  This lack of increase for $\alpha_{\rm vir}$ does not alone imply that feedback does not increase the molecular velocity dispersion, because feedback could agitate the GMCs which then expand a little as they absorb the additional energy, while maintaining their near-virial signature. However, GMCs cannot satisfy both the Larson Laws, which include the condition $\alpha_{\rm vir}\sim$constant, and increase their velocity dispersions or radii without increasing their masses, and star formation does not increase a cloud's mass. We return to this point also in Section \ref{total} where the slopes of the Larson power-law relations in our data are shown to be the same as the slopes of the excursion relations for azimuthal variations.

Figure \ref{n4321-deltas} shows a clear difference in two panels between the values derived from individual, deconvolved clouds (red points) and value derived inside 150 pc pixels regardless of whether they center on a cloud or not (blue points). Because of this difference, we maintain this distinction throughout the paper. The difference manifests as a steeper slope for $\Delta\Sigma_{\rm mol}(\Delta\Sigma_{\rm SFR})$ and a shallower slope for $\Delta\sigma(\Delta\Sigma_{\rm mol})$ for the objects than the pixels. This is probably because $\Sigma_{\rm mol}$ for the pixels is averaged over cloud and intercloud emissions inside each 150 pc region, giving it a smaller range than for $\Sigma_{\rm mol}$ peaked-up on each cloud.

\begin{figure}
\includegraphics[width=12cm]{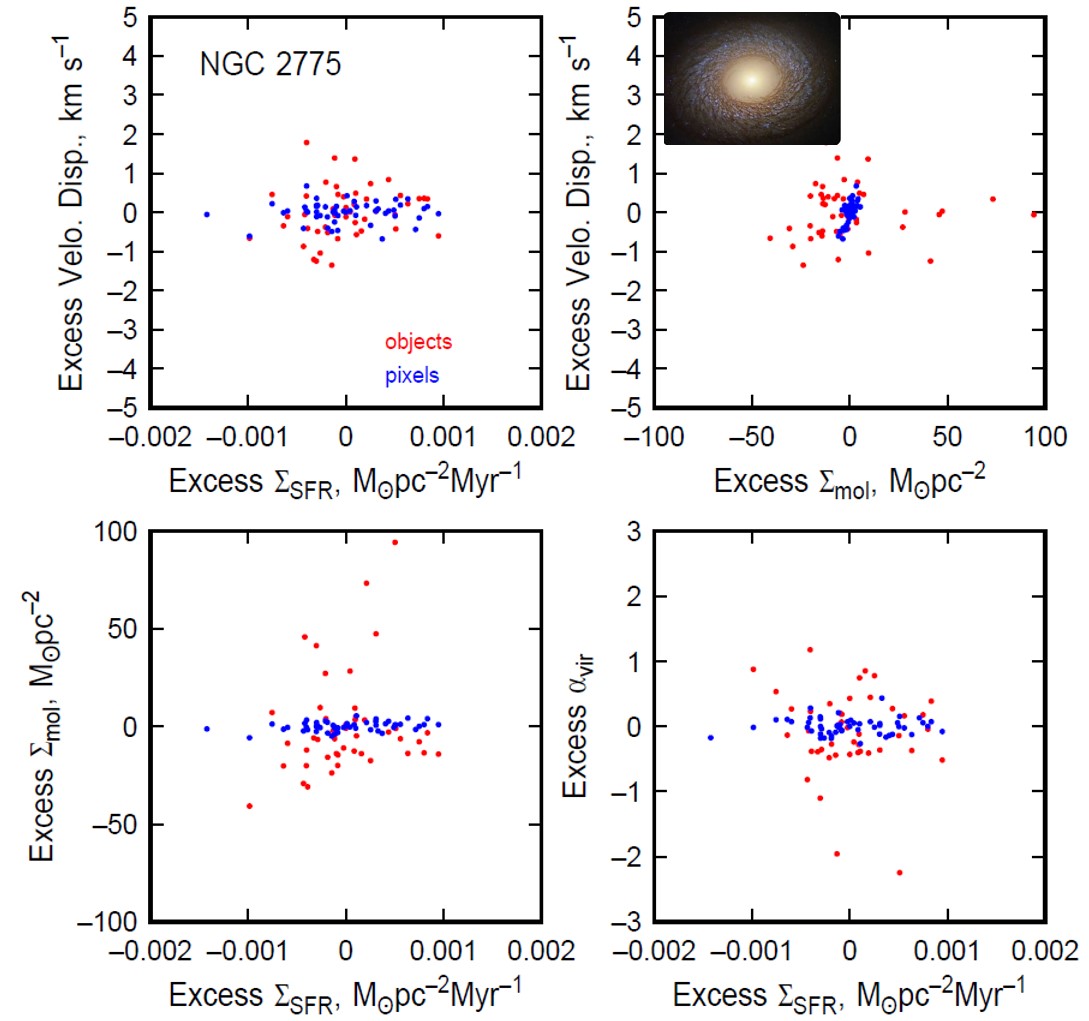}
\caption{Correlations between azimuthal fluctuations for the weak-arm galaxy NGC 2775. The ranges for $\Sigma_{\rm SFR}$ and $\Sigma_{\rm mol}$ are approximately one-tenth the ranges for NGC 4321, as reflected in the smaller scales on the figure axes here. Symbol colors are as in Fig. \ref{n4321-deltas}.}
\label{n2775-deltas}
\centering
\end{figure}

The large variations in $\Delta\Sigma_{\rm SFR}$ in Figure \ref{n4321-deltas}, amounting to a variation from $-0.005$ to +0.02$ \;M_\odot$ pc$^{-2}$ Myr$^{-1}$, are mostly the result of azimuthal variations in the SFR and molecular cloud density with the spiral phase: both are much higher in the arms. This is in contrast to the case for a flocculent galaxy, NGC  2775, where the PHANGS values are shown in Figure \ref{n2775-deltas}. Note that the scale for $\Delta \Sigma_{\rm SFR}$ is blown-up by a factor of 10 compared to Figure \ref{n4321-deltas}, and the scale for $\Delta\Sigma_{\rm mol}$ is blown up by a factor of 2. This difference is sensible because flocculent galaxies have much more uniform SFRs around in azimuth than galaxies with strong spiral density waves like NGC  4321, where the star formation is concentrated in the arms.  In what follows, the strong-arm and weak arm galaxies will be distinguished by different colored symbols or different figures. 

Now we consider a larger sample of galaxies from the PHANGS survey. We examined plots like these for all of the galaxies in release 4 at the link given above and found that galaxies with stellar masses less than around $2\times10^{10}\;M_\odot$ had too few CO observations to make good correlations. We also found that several galaxies with larger masses than this had too few CO points also. These were eliminated from further study here. Table \ref{table:sample} lists all three classes of galaxies: strong-arm with a relatively large range in $\Delta\Sigma_{\rm SFR}$, weak-arm or flocculent galaxies with a relatively small range in $\Delta\Sigma_{\rm SFR}$, and galaxies with too few CO observations to be useful here, all with $M_{\rm star}>2\times10^{10}\;M_\odot$. 

Figure \ref{all-pix} shows the same quantities as in Figures \ref{n4321-deltas} and \ref{n2775-deltas} but now for the full sample using pixel values, distinguishing broad-range $\Delta \Sigma_{\rm SFR}$ with blue symbols and narrow-range $\Delta \Sigma_{\rm SFR}$ with red symbols. The differences between these galaxy types in all the panels are clear, but within each panel, there are tight correlations for each one. Best fit correlations are shown by the lines, as determined with bivariable linear fits because there is a dispersion along each axis. We limit each fit to the ranges of variables shown in the figure, as some galaxies, particularly near their nuclei, have much larger variations in some quantities that are far off the scales here. (In a bivariable fit, we first fit $y=A_1+B_1x$ for abscissa $x$ and ordinate $y$ values and then fit $x=A_2+B_2y$ for the other direction. The bivariate fit is $y=A_3+B_3x$ for $A_3=(A_1-A_2/B_2)/2\pm(A1+A2/B2)/2$ and $B_3=(B_1+1/B_2)/2\pm(B1-1/B2)/2)$.) The fits are given in Table \ref{table:linearfits2}. With this type of fitting, the inverse correlation, $B(A)$, has a slope equal to the inverse of the direct correlation, $A(B)$. 

Figure \ref{all-obj} shows the same parameters for the cloud-object measurements, with bivariate linear fits also in Table \ref{table:linearfits2}. The pixel and object correlations are about the same in the two figures, and for each, the difference between broad-range $\Sigma_{\rm SFR}$ (strong spiral arms, blue symbols) and narrow range $\Sigma_{\rm SFR}$ (weak arms or flocculent structure with red symbols) are consistent: the weak-arm correlations are steeper. 

We show now that the correlation between excess molecular velocity dispersion and excess star formation rate density is trivially related to two other correlations: one between excess velocity dispersion and excess molecular mass and another between excess molecular mass and excess star formation. Writing 
\begin{equation}
\Delta\sigma=A_{\sigma,{\rm mol}}+B_{\sigma,{\rm mol}}\Delta\Sigma_{\rm mol}
\end{equation}
and
\begin{equation}
\Delta\Sigma_{\rm mol}=A_{\rm mol,SFR}+B_{\rm mol,SFR}\Delta\Sigma_{\rm SFR}
\end{equation}
from the fits in Table \ref{table:linearfits2}, we expect the coefficients in the third correlation,
\begin{equation}
\Delta\sigma=A_{\sigma,{\rm SFR}}^\prime+B_{\sigma,{\rm SFR}}^\prime\Delta\Sigma_{\rm SFR}
\end{equation}
to be 
\begin{equation}
A_{\sigma,{\rm SFR}}^\prime=A_{\sigma,{\rm mol}}+B_{\sigma,{\rm mol}}A_{\rm mol,SFR}
\end{equation}
and
\begin{equation}
B_{\sigma,{\rm SFR}}^\prime=B_{\sigma,{\rm mol}}B_{\rm mol,SFR}.
\end{equation}
Table \ref{table:linearfits2} gives these values also and shows they compare well with the directly measured correlation between $\Delta \sigma$ and $\Delta\Sigma_{\rm SFR}$. The correspondence is not exact because of our fitting limits to the ranges in Figures \ref{all-pix} and \ref{all-obj}.

Because star formation follows molecular mass, it is most likely that the $\Delta\Sigma_{\rm SFR}(\Delta\Sigma_{\rm mol})$ correlation is primary, or causal, i.e., star formation depends on the molecular gas. In that case, because more mass corresponds to higher turbulent speeds through intrinsic molecular cloud correlations \citep[the ``Larson Laws'';][]{larson81}, it is likely that the $\Delta \sigma(\Delta\Sigma_{\rm SFR})$ correlation is secondary and not causal. If, conversely, larger $\sigma$ were the result of feedback with a locally larger SFR, then there would have to be an unphysical addition of molecular mass after star formation begins to increase $\Sigma_{\rm mol}$ up to the correlated value with $\sigma$ (considering that the virial parameter is about constant). 

\section{Total parameter variations in the PHANGS survey}
\label{total}

We now look at the full values of $\sigma$, $\Sigma_{\rm mol}$, $\Sigma_{\rm SFR}$ and $\alpha_{\rm vir}$ in our sample galaxies, without subtracting the average radial profiles. These are first shown in Figure \ref{n4321-full} for NGC 4321 with red and blue symbols again distinguishing between object measurements and pixel measurements. The correlations for these two measurement types are about the same although the object values have slightly higher $\sigma$ and $\Sigma_{\rm mol}$, and slightly lower $\alpha_{\rm vir}$ than the pixel values, again most likely because of the averaging process in the pixel measurement. 

\begin{figure}
\includegraphics[width=12cm]{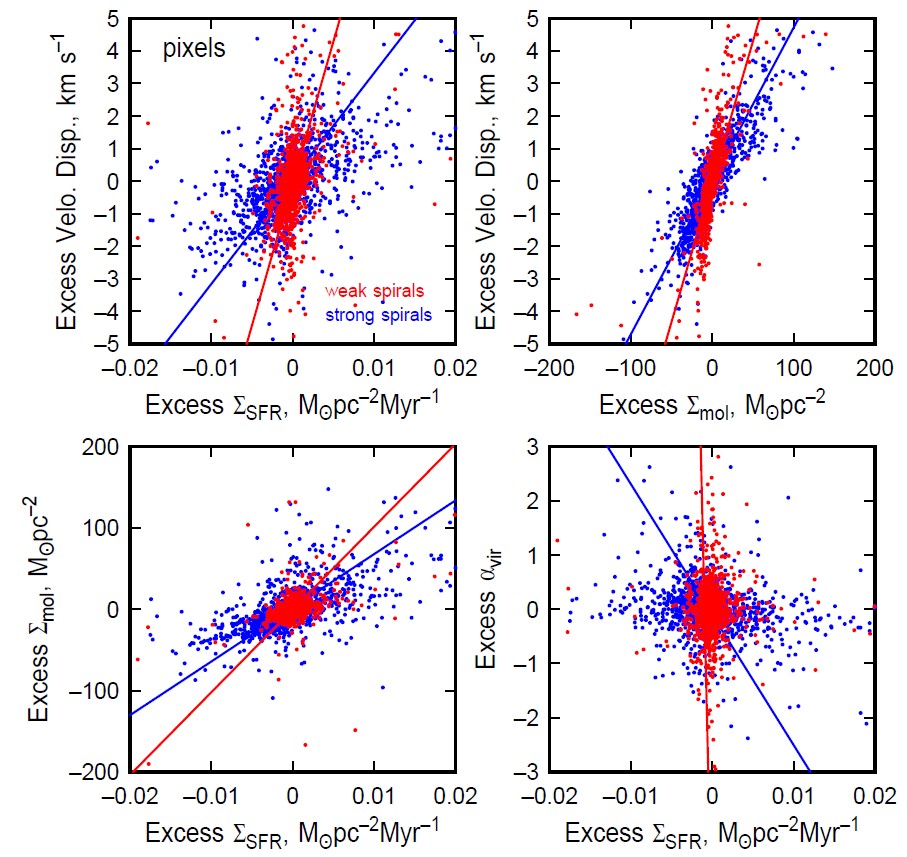}
\caption{Correlations between azimuthal fluctuations of various quantities for 34 galaxies in the PHANGS survey with stellar masses larger than $2\times10^{10}\;M_\odot$. This figure uses data measured in regular pixels 150 pc in diameter. Red and blue lines are fits to the trends corresponding to the red and blue points, which are for weak and strong spirals, respectively. The fitting results are in Table \ref{table:linearfits2}.}
\label{all-pix}
\centering
\end{figure}

\begin{figure}
\includegraphics[width=12cm]{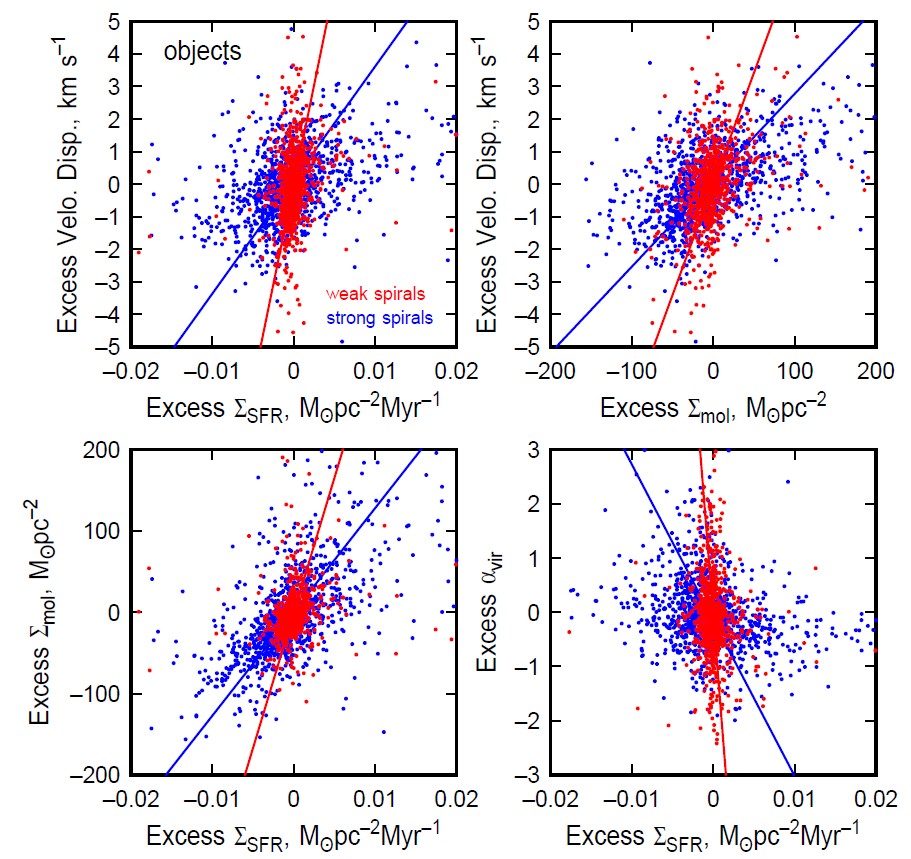}
\caption{As in Figure \ref{all-pix} but with data measured in 150 pc regions for molecular cloud objects and averaged together in 1.5 kpc hexagons.  Red and blue lines are fits to the trends corresponding to the red and blue points, which are for weak and strong spirals, respectively. The fitting results are in Table \ref{table:linearfits2}.}
\label{all-obj}
\centering
\end{figure}

\begin{figure}
\includegraphics[width=12cm]{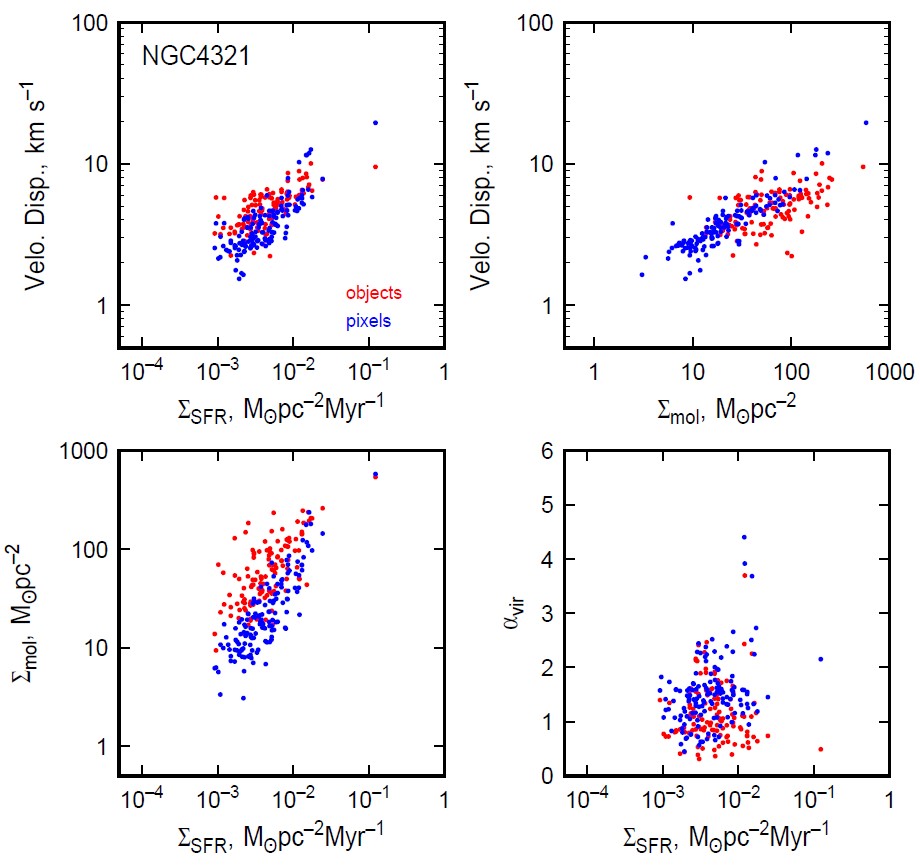}
\caption{Correlations between total quantities in NGC 4321, plotted on a log-log scale.  The red points are from object measurements and the blue points are from pixel measurements, both at 150 pc resolution and average into 1.5 kpc hexagons. }
\label{n4321-full}
\centering
\end{figure}

\begin{figure}
\includegraphics[width=12cm]{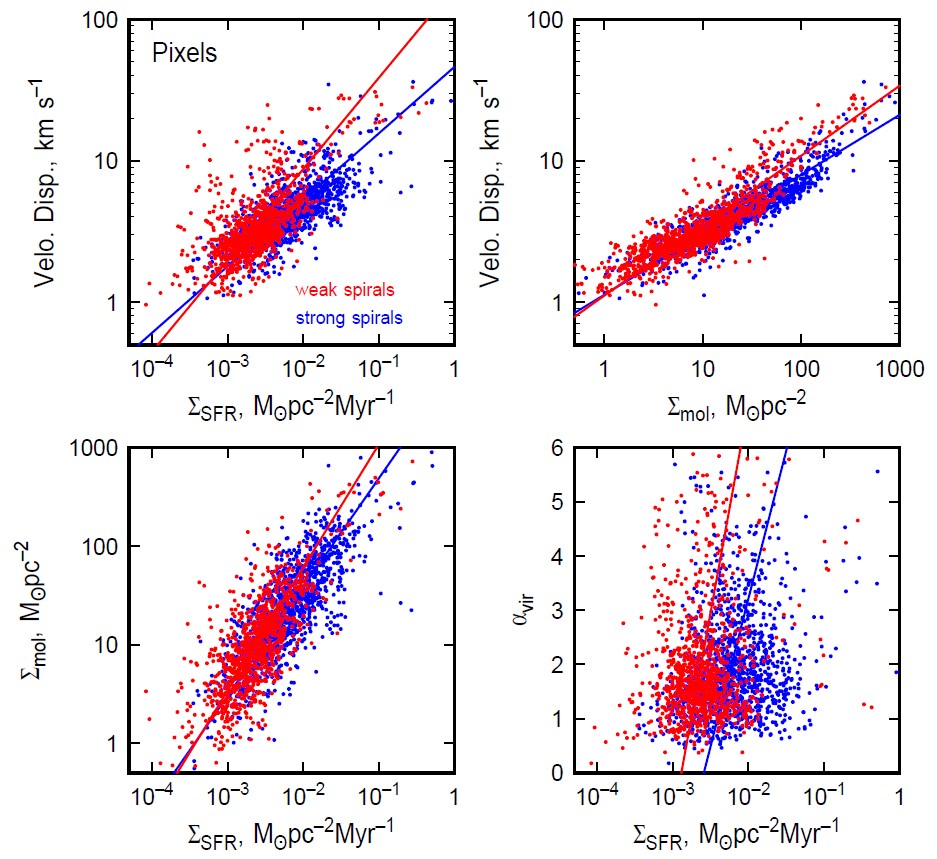}
\caption{Correlations between total quantities in all 34 galaxies considered here, using pixel measurements with 150 pc resolution. Weak and strong spirals are distinguished by color, with lines showing the linear fits in log-log space. The fitting results are in Tables \ref{table:linearfits2} and \ref{table:powerlawfits}.}
\label{all-full-pix}
\centering
\end{figure}

\begin{figure}
\includegraphics[width=12cm]{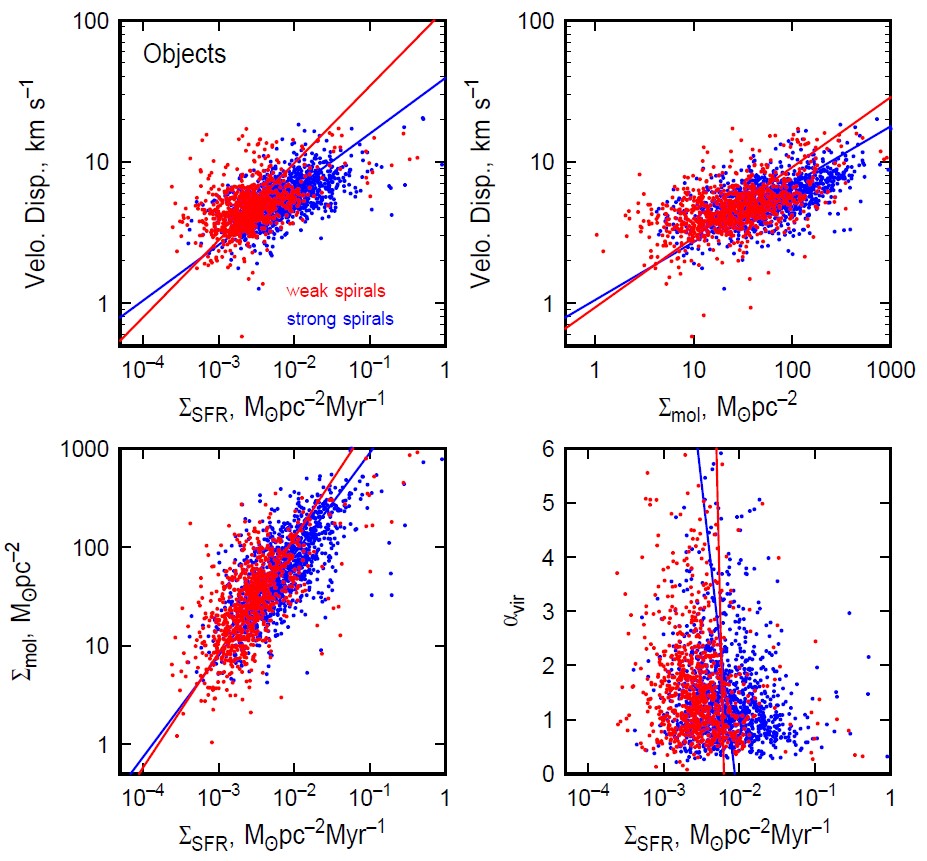}
\caption{Correlations between total quantities using object measurements with 150 pc resolution. Weak and strong spirals are distinguished by color, with lines showing the linear fits in log-log space. The fitting results are in Tables \ref{table:linearfits2} and \ref{table:powerlawfits}.}
\label{all-full-obj}
\centering
\end{figure}

Figure \ref{all-full-pix} shows the same quantities for the pixel measurements in our full survey, distinguishing between strong spiral-arm (blue symbols) and weak spiral-arm (red) galaxies.  Figure \ref{all-full-obj} shows the object measurements. In each figure, the full parameter correlations for molecular clouds are about the same for the two galaxy types, which means that the Larson Law, $\sigma(\Sigma_{\rm mol})$ (top right), is nearly independent of spiral arms, as is the efficiency of star formation per unit CO mass (lower left). The weak-arm galaxies are systematically shifted toward lower $\Sigma_{\rm SFR}$ and $\Sigma_{\rm mol}$, in addition to having a lower range of these values with azimuth (cf. Figs. \ref{all-pix} and \ref{all-obj}). These shifts in full values are presumably because either the total star formation rates in weak-arm galaxies are smaller than in strong-arm galaxies, or because the star formation is more spread-out in weak-arm galaxies with about the same total rate (depending on the galaxy). 

The coefficients $C$ and slopes $D$ of the power law correlations in Figures \ref{all-full-pix} and \ref{all-full-obj} are given in Table \ref{table:powerlawfits}, considering equations like $\log y=C + D\log x$. These are from bivariate fits in log-log space except for $\alpha_{\rm vir}(\Sigma_{\rm SFR})$, which is in linear-log space, keeping $\alpha_{\rm vir}$ linear as in the figures. 

\begin{figure}
\includegraphics[width=12cm]{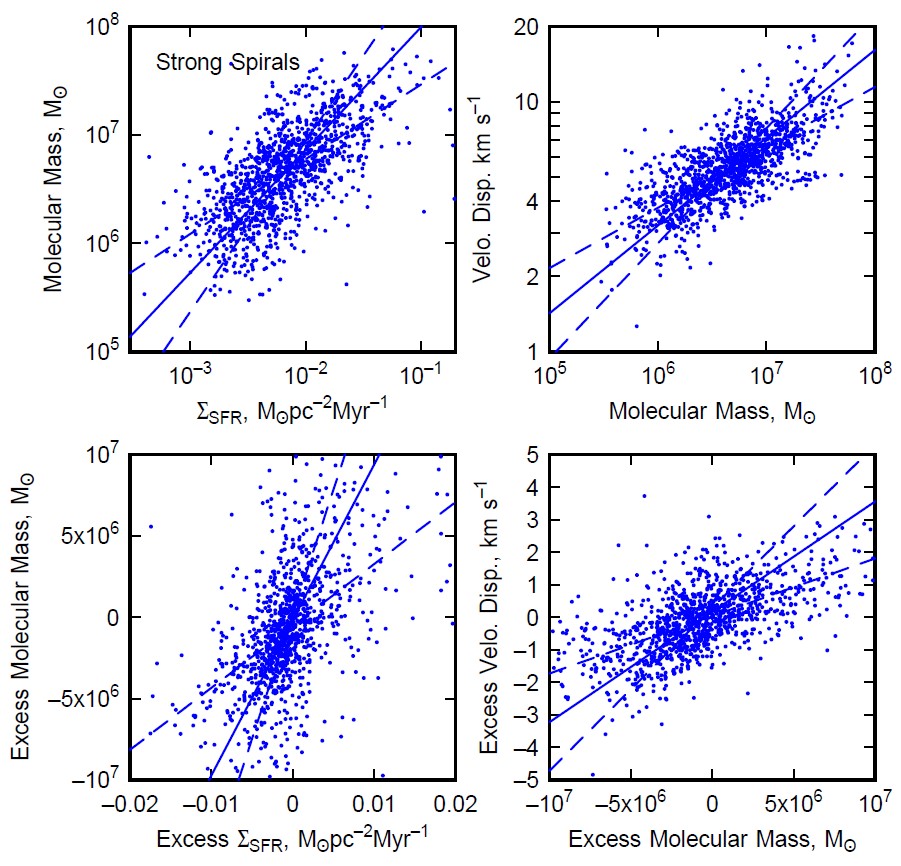}
\caption{Correlations between azimuthal fluctuations involving molecular cloud mass. This figure is for the 17 strong-arm galaxies in our survey. Fitting parameters are in Table \ref{table:powerlawfits}.}
\label{broad-full-mass}
\centering
\end{figure}

\begin{figure}
\includegraphics[width=12cm]{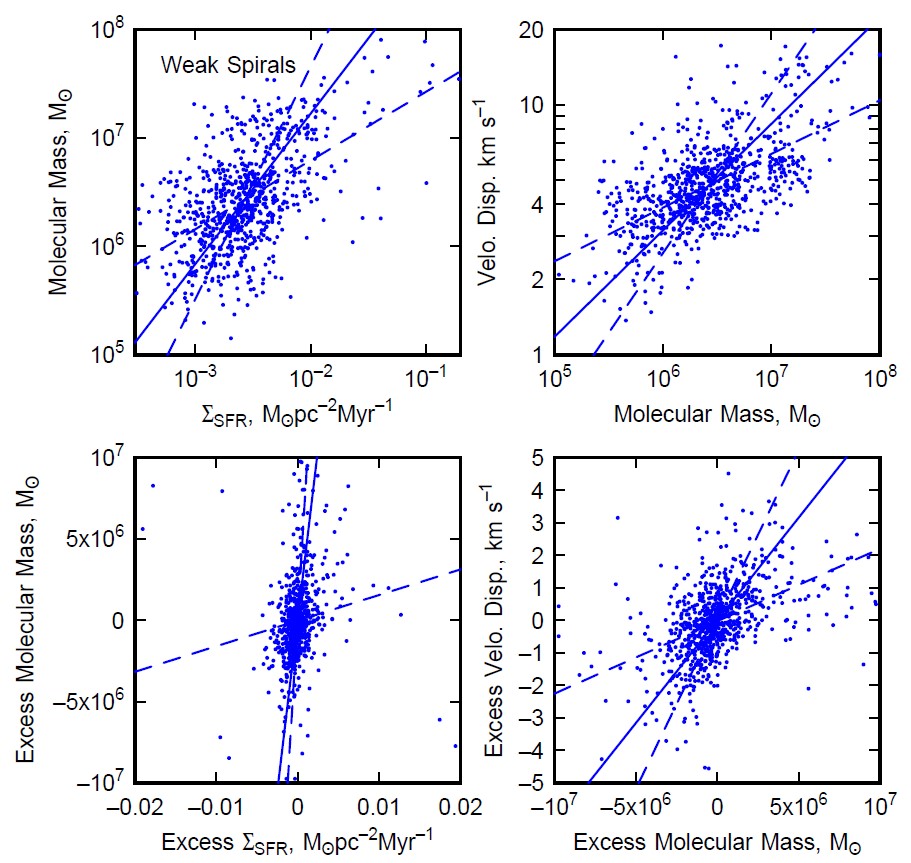}
\caption{Correlations between azimuthal fluctuations involving molecular cloud mass. This figure is for the 17 weak-arm galaxies in our survey. Fitting parameters are in Table \ref{table:powerlawfits}.}
\label{narrow-full-mass}
\centering
\end{figure}

Now we ask whether the excursions about the azimuthal averages, shown in Figures \ref{all-pix} and \ref{all-obj}, are the result of more or less CO mass inside each measured region, following the full correlations seen in Figures \ref{all-full-pix} and \ref{all-full-obj}. We do this for both the pixel measurements, which have $\Sigma_{\rm mol}$ for molecular cloud content in a pixel, and the object measurements, which have both $\Sigma_{\rm mol}$ and average cloud mass, $M_{\rm mol}$, in each measured region. Because we have not mentioned $M_{\rm mol}$ yet, Figures \ref{broad-full-mass} and \ref{narrow-full-mass} show how the average molecular cloud mass and the azimuthal variations of this average molecular mass correlate with $\sigma$ and $\Sigma_{\rm SFR}$ for strong-arm and weak-arm galaxies, respectively. The top right in each figure, $\sigma(M_{\rm mol})$, is another standard molecular cloud correlation (Larson Law). The top left depends on the efficiency of star formation per molecular cloud (although the abscissa is per unit area and the ordinate is not). Fits to the power laws involving cloud mass are also in Tables 2 and 3. The dashed lines in Figure \ref{narrow-full-mass} are error limits to the slopes (not shown in the other figures for clarity). 

To determine if the full correlations are entirely responsible for the azimuthal variations (which is a non-trivial result), we determine the slopes of the full correlations and compare them to the ratios of excursions in the azimuthal variations. That is, we examine whether the power-law relations for total quantities in Figures \ref{all-full-pix} and \ref{all-full-obj}, such as
\begin{equation}
\sigma=10^{C_{\sigma,{\rm SFR}}}\Sigma_{\rm SFR}^{D_{\sigma,{\rm SFR}}},
\end{equation}
produce the linear relations of the excursion quantities in Figures \ref{all-pix} and \ref{all-obj}, correspondingly,
\begin{equation}
\Delta \sigma=A_{\sigma,{\rm SFR}}+B_{\sigma,{\rm SFR}}\Delta\Sigma_{\rm SFR}.
\end{equation}
For this to be true for the slope (since the intercept cancels out in the excursion relations), we need
\begin{equation}
B_{\sigma,{\rm SFR}}^\prime={D_{\sigma,{\rm SFR}}}\left({{<\sigma>}\over{<\Sigma_{\rm SFR}>}}\right)
\label{eq:checkslope}
\end{equation}
to equal the measured slope of the excursion, $B_{\sigma,{\rm SFR}}$,
where the brackets, $<>$, denote averages. Equation \ref{eq:checkslope} comes from the derivative of the power law, setting $\Delta\sigma/\Delta\Sigma_{\rm SFR}=d\sigma/d\Sigma_{\rm SFR}$ at the average value of each quantity.

Table \ref{table:comparisons} compares the slope $B_{\sigma,{\rm SFR}}$ of the correlation between excess velocity dispersion and excess SFR with the slope $B^\prime_{\sigma,{\rm SFR}}$ of the correlation between total dispersion and SFR from the power laws. The average values used for equation \ref{eq:checkslope} are from the averages of the logs of these quantities, since the logs were used in the fit to the power law. The second and sixth columns indicate that $B_{\sigma,{\rm SFR}}^\prime\sim B_{\sigma,{\rm SFR}}$. This result suggests that the local excess in $\sigma$ in star-forming regions is the result of the intrinsic properties of molecular clouds, which have normal power-law relationships in our galaxies. Because these properties are established prior to and during star formation, the excess $\sigma$ in a region of star formation cannot be entirely the result of feedback from that star formation.

\section{Correlations with Pressure}
\label{sect:pressure}
\subsection{General Pressure Trends}
\label{generalP}
Having examined correlations between molecular cloud turbulence through the velocity dispersion $\sigma$ and the local SFR, we turn now to correlations between local pressure $P$ and SFR. If there is significant feedback, then one might expect $P$ to increase with SFR, especially for the older regions where feedback has been working for a long time. However, $P$ also increases with $\Sigma_{\rm mol}$ and $\sigma$, which, as we have seen, increase in a region as molecular clouds form before star formation begins, so a correlation like $\Delta P(\Delta\Sigma_{\rm SFR}$) could also be secondary. 

Figure \ref{pressure} shows various correlations between pressure and SFR, with azimuthal variations in the bottom panels on a linear scale and total parameter values in the upper panels on a log scale. We consider three measures of pressure: $P_{\rm P150}$ is the molecular pressure for 150 pc pixels in the PHANGS tabulation, averaged over 1.5 kpc hexagons; $P_{\rm O150}$ is the molecular pressure for cloud objects in 150 pc regions, averaged over 1.5 kpc hexagons (both from the PHANGS tabulation), and $P_{\rm DE}$ is the dynamical equilibrium pressure from \cite{ostriker22}, written in equation 1 of \cite{sun23}:
\begin{equation}
P_{\rm DE}={{\pi G}\over{2}}\Sigma_{\rm gas}^2 + \Sigma_{\rm gas}\left(2G\rho_{\rm star}\right)^{1/2}\sigma_{\rm gas,z}
\label{eq:pde}
\end{equation}
where $\Sigma_{\rm gas}=\Sigma_{\rm mol}+\Sigma_{\rm atom}$ is the total gas surface density from molecules and atoms, $\rho_{\rm star}$ is the midplane stellar density, and $\sigma_{\rm gas,z}$ is the effective vertical component of the gas velocity dispersion, assumed by \cite{sun23} and also here to be a constant $11$ km s$^{-1}$ (including magnetic and cosmic-ray pressures).  All of these pressures are uncertain because of resolution limits (e.g., cloud pressure depends on density which depends on radius) and unknown quantities, such as $\rho_{\rm star}$ and $\sigma_{\rm gas,z}$ in $P_{\rm DE}$. Also, the nature of molecular gas motions is not known, e.g., whether they are turbulent or collapsing, compressional or rotational, etc., and a wide combination of motions could give $\alpha_{\rm vir}\sim1$ when self-gravity is important. 

The bivariate fits to the three $P(\Sigma_{\rm SFR})$ correlations are in Table \ref{table:pressure}. The excursion correlations have linear fits in the second and third columns, and the total correlations have power law fits in the fourth and fifth columns. 

There are several points to make from Figure \ref{pressure} and Table \ref{table:pressure}. First, the correlations between molecular cloud pressures ($P_{\rm P150}$ and $P_{\rm O150}$) and pressure excursions ($\Delta$-values) with $\Sigma_{\rm SFR}$ and its excursions are steeper for weak spirals than strong spirals.  For example, the weak-spiral cloud objects have a $P(\Sigma_{\rm SFR})$ correlation with a slope of $\sim16900$ in the table (in units of $10^4$ K cm$^{-3}/(M_\odot$ pc$^{-2}$ Myr$^{-1})$), while for strong spirals the slope is $\sim7210$, a factor of 2.3 smaller. For molecular pixels, the ratio between the two slopes is $\sim15500/5730=2.7$. Similarly for the power-law relations for total quantities, the powers are $\sim1.92$ and $\sim1.59$ for cloud objects in weak and strong and spirals, respectively, and $\sim2.5$ and $\sim2.0$ for pixel measurements in weak and strong spirals. 

The reason for this arm strength difference is not apparent. Another way of describing it is that a given increase in $\Sigma_{\rm SFR}$ has a greater impact on $P_{\rm P150}$ and $P_{\rm P150}$ in a weak spiral arm than a strong spiral arm, or, alternatively, the molecular cloud pressure has a greater impact on $\Sigma_{\rm SFR}$ in strong arms than weak arms. In this latter interpretation, the arm would be doing something extra to the gas to stimulate additional star formation. That is, stronger spiral arms may have a higher efficiency of star formation in a given molecular cloud than weaker spiral arms. This is a sensible interpretation, but it implies that molecular cloud variations, including the pressure variations that accompany different molecular clouds, are driving the azimuthal star formation variations, rather than that star formation is pressurizing, more or less, the molecular clouds. 

A second point is that azimuthal variations in molecular cloud pressure, $\Delta P_{\rm P150}$ and $\Delta P_{\rm O150}$, are much larger in ratio to the SFR variations than the derivative of the total cloud pressures with $\Sigma_{\rm SFR}$. To see this in Table \ref{table:pressure}, compare the $B$ values to the $B^\prime$ values just below them for the O150 and P150 pressures; i.e., $B >> B^\prime$. This result differs from the $\Delta\sigma(\Delta\Sigma_{\rm SFR})$ correlations discussed in the previous section, which always followed from the power law relations for the total quantities, $\sigma(\Sigma_{\rm SFR})$.  Perhaps the reason for the difference in pressure correlations is that the total pressure variations in the power law contain a radial dependence and a galaxy-to-galaxy variation, which would introduce significant variations in the background stellar gravity. Then the derivative of total pressure with SFR density is diluted by the background stars and becomes smaller than the pressure fluctuations in azimuth, which are nearly pure gas. 

A third point is that the spiral-strength differences in cloud pressure mentioned above contrast with the dynamical pressure, $P_{\rm DE}$, and its azimuthal variations $\Delta P_{\rm DE}$.  The power-law slope of $P_{\rm DE}(\Sigma_{\rm SFR})$ is about the same in both spiral arm types (1.25 and 1.33). Also, the ratio of excursion pressure to excursion SFR is about equal to the predicted value from the slope of the total-pressure power law: 979 compared to 909 for strong spirals and 882 compared to 1180 for weak spirals.  These results imply that dynamical pressure and SFR follow each other closely for both azimuthal and radial variations and from galaxy to galaxy, unlike the molecular pressures.

Fourth, the azimuthal variations in dynamical pressure are $\sim10$ times smaller than the azimuthal variations in molecular cloud pressure (compare the y-axis limits in the lower right of Figure \ref{pressure} with the limits in the lower left and center). Thus, the slope of the $\Delta P_{\rm DE}(\Delta\Sigma_{\rm SFR})$ correlation is $\sim10$ times smaller than the slopes of the molecular cloud correlations. This is true even though the total pressure ranges are about the same (top panels). This result suggests that molecular clouds come and go in regions of more or less star formation, and the molecular cloud pressures have large positive and negative excursions in these regions, but the dynamical pressure around them hardly changes. This difference could again be the result of the disk stellar contribution to the dynamical pressure, whereas molecular cloud pressures contain substantial self-gravity. 

Fifth, the molecular cloud pressures are 10 or more times larger than the dynamical pressures at high $\Sigma_{\rm SFR}$ (comparing the top three panels in Fig. \ref{pressure}). We can quantify this from the fits to the power laws in Table \ref{table:pressure}. At the high-end of the SFR, $\Sigma_{\rm SFR}=0.1\;M_\odot$ pc$^{-2}$ Myr$^{-1}$, the three pressures are, on average: $P_{\rm O150}\sim 1.05\times10^3$, $P_{\rm P150}\sim 1.58\times10^3$ and $P_{\rm DE}\sim 0.145\times10^3$ for strong-arm spirals, in units of $10^4$ K cm$^{-3}$. These numbers are $5.89\times10^3$, $22.4\times10^3$ and $0.295\times10^3$ for the weak-arm spirals. Clearly, $P_{\rm O150}$ and $P_{\rm P150}$ are much larger than $P_{\rm DE}$. At the low-end of the SFR, where $\Sigma_{\rm SFR}=10^{-3}\;M_\odot$ pc$^{-2}$ Myr$^{-1}$, all of these pressures are about the same, ranging from 0.2 to 0.9 in units of $10^4$ K cm$^{-3}$. These differences again are explained if molecular cloud pressure is dominated by self-gravity. 

Sixth, the molecular cloud pressures increase with $\Delta\Sigma_{\rm SFR}$, unlike what we would expect if feedback energy puffs up a GMC at fixed mass and gives it a lower surface density, which corresponds to a lower GMC pressure and a lower virialized velocity dispersion.  This observation suggests again that GMC properties, including pressure and velocity dispersion, are precursors to star formation tied to the cloud mass, and that $\Sigma_{\rm SFR}$ follows the cloud mass too.

In summary, the molecular pressure is up to 10 times higher than the dynamical pressure everywhere, and it is $\sim10$ times more sensitive to local $\Sigma_{\rm SFR}$ variations than the dynamical pressure. This is probably because molecular clouds are at higher pressures than the average ISM as a result of their strong internal gravities. This is consistent with virial parameters of order unity in star-forming clouds. If their boundary pressures were larger than their self-gravitating pressures, then $\alpha_{\rm vir}$ would be much larger than 1. 

\subsection{Reflections on $P_{\rm DE}$: Feedback or Precursor to Star Formation?}
\label{reflect}

\begin{figure}
\includegraphics[width=16cm]{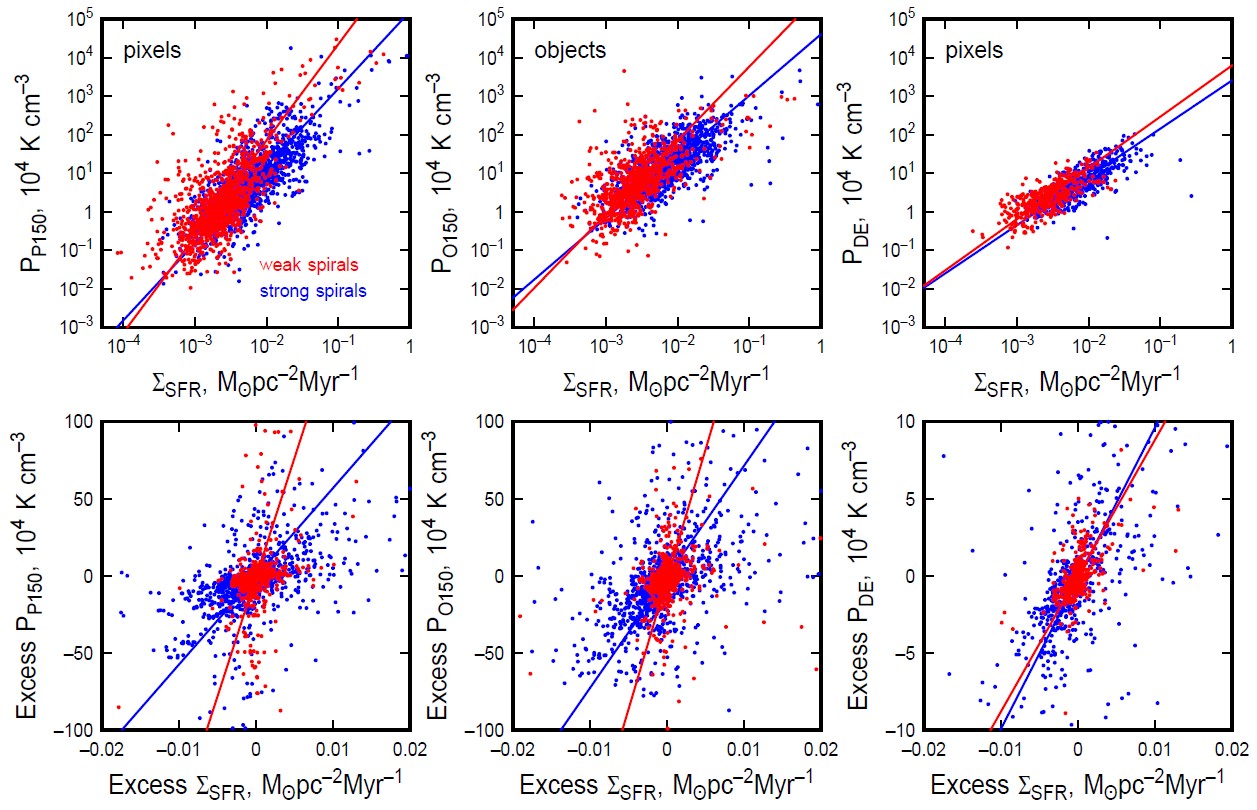}
\caption{(bottom) Correlations between azimuthal fluctuations in $\Sigma_{\rm SFR}$ and azimuthal fluctuations in three measures of pressure: two for molecular clouds using the object- and pixel-based methods, and one for the dynamical equilibrium pressure, $P_{\rm DE}$. (top) Correlations between the total quantities represented by their fluctuations in the bottom panels. Red and blue points and fitting lines are for galaxies with weak and strong spiral arms. Fitting parameters are in Table \ref{table:pressure}.  }
\label{pressure}
\centering
\end{figure}

The trend for $P_{\rm DE}$ in Figure \ref{pressure} may be compared with expectations from previous work. The slopes of the $P_{\rm DE}(\Sigma_{\rm SFR})$ correlations found here, namely, $1.25\pm0.30$ for strong spirals and $1.33\pm0.40$ for weak spirals, differ from the slopes in \cite{ostriker22} and \cite{sun23}. \cite{ostriker22} considered the inverse relationship and obtained from simulations $\log\Sigma_{\rm SFR}=1.21\log(P_{\rm DE}/k_{\rm B})-7.66$ for $\Sigma_{\rm SFR}$ in units of $M_\odot$ pc$^{-2}$ Myr$^{-1}$.  Rearranging terms, that would be $\log(P_{\rm DE}/k_{\rm B})=C+D\log(\Sigma_{\rm SFR})$ where $C=7.66/1.21=6.33$ and $D=1/1.21=0.83$.  In table \ref{table:pressure}, we get $C=7.41$ (converting here to pressure in $k_{\rm B}$ instead of $10^4k_{\rm B}$ in the table) and $D=1.25$ for strong spirals and $C=7.80$, $D=1.33$ for weak spirals. Although the fitted slopes differ (0.83 compared to 1.3 here), all of these values give about the same $P_{\rm DE}$ for a typical $\Sigma_{\rm SFR}=0.01\;M_\odot$ pc$^{-2}$ Myr$^{-1}$; i.e., \cite{ostriker22} would get $\log(P_{\rm DE}/k_{\rm B})=4.67$ from their fit, and we get $\log(P_{\rm DE}/k_{\rm B})=4.91$ and $5.14$ for strong and weak spirals.  In their comparison with observations (Fig. 15) forcing a unit slope, \cite{ostriker22} suggest $\Sigma_{\rm SFR}=2.07\times10^{-7}P_{\rm DE}$, which gives $\log(P_{\rm DE}/k_{\rm B})=4.68$ at $\Sigma_{\rm SFR}=0.01$ for their choice of feedback yield. Actually, the observations they plot appear to have a shallower slope than 1.21 for $\Sigma_{\rm SFR}(P_{\rm DE})$, more like unity, which brings it closer to the slope in Figure \ref{pressure} here.

Equation \ref{eq:pde} contains several assumptions, so the value of the slope could change with different assumptions. We used a star formation rate from FUV and WISE $22\mu$ emission, and as a check, redid all the calculations using $H\alpha$ plus WISE $22\mu$ (both tabulated in release 4 of PHANGS data). The resultant slope for $P_{\rm DE}(\Sigma_{\rm SFR})$ differed only in the 3rd decimal place, which is a negligible difference compared to the uncertainty. Also uncertain is the CO to mass conversion factor, $\alpha_{\rm CO}$, but the one we use is not much different than others as long as reasonable variations with metallicity, velocity dispersion or other co-varying parameters are considered. 

More important to the $P_{\rm DE}(\Sigma_{\rm SFR})$ slope is the assumption about stellar scale height, $H_{\rm star}$, which enters into the stellar density $\rho_{\rm star}$ during conversion from the stellar surface density. \cite{ostriker22} use a constant $H_{\rm star}$ in each of two sets of simulations, while the PHANGS data we use assumes $H_{\rm star}$ increases with galactocentric radius as in Figure 6 of \cite{sun20}. Going from a constant $H_{\rm star}$ to an increasing $H_{\rm star}$ with radius lowers $P_{\rm DE}$ at low $\Sigma_{\rm SFR}$, steepening the $P_{\rm DE}(\Sigma_{\rm SFR})$ slope.  This makes the inverse \cite{ostriker22} slope of 0.83 (see above) larger (in the direction of our slope) if variable $H_{\rm star}$ were used by them. \cite{sun23} fit PHANGS data to a slope for $d\Sigma_{\rm SFR}/dP_{\rm DE}$ equal to 0.93 in the fiducial case, which corresponds to 1.07 for the inverse correlation in our Figure \ref{pressure}.  This is closer to our value of $\sim1.3$, and \cite{sun23} assume H$_{\rm star}$ increases with radius as well. We note also that our slope of $1.3$ for $P_{\rm DE}(\Sigma_{\rm SFR})$ is similar to that found by \cite{fisher19} (i.e., $1/0.75=1.3$) for \HI\ in redshift $z\sim1$ turbulent disks.  

A constant effective gaseous vertical velocity dispersion was also assumed for equation \ref{eq:pde}. This effective dispersion is from a combination of thermal and turbulent motions, with magnetic and cosmic ray contributions. If $\sigma_{\rm gas,z}$ systematically decreases with radius instead, which seems reasonable as observed turbulent speeds tend to do this \citep{stilp13}, that would decrease $P_{\rm DE}$ in the outer regions where $\Sigma_{\rm SFR}$ is low, and increase the slope of $P_{\rm DE}(\Sigma_{\rm SFR})$ even more, taking it further from unity.

Because of these uncertainties and differing assumptions, we do not consider the slope of $P_{\rm DE}(\Sigma_{\rm SFR})$ to be significantly different from other values in the literature. Nevertheless, we note that if the slope is systematically different from unity, then the feedback yield from star formation, $P_{\rm DE}/\Sigma_{\rm SFR}$ \citep{ostriker22}, would have to vary with pressure or SFR in about the same way for both strong and weak arm galaxies (Fig. \ref{pressure} top-right), and also as these quantities vary around in azimuth in any one galaxy (Fig. \ref{pressure} bottom-right), considering the slopes of $P_{\rm DE}(\Sigma_{\rm SFR})$  and $\Delta P_{\rm DE}(\Delta \Sigma_{\rm SFR})$ are the same (Table \ref{table:pressure}). 

Equation \ref{eq:pde} contains two terms that each resemble a Kennicutt-Schmidt relation. The first term depends only on $\Sigma_{\rm gas}$, like the original KS relation \citep{kennicutt12}, and the second term depends only on $\Sigma_{\rm gas}(\rho_{\rm star})^{1/2}$ (because $\sigma_{\rm gas,z}$ is assumed to be constant), which is analogous to the Extended KS relation discussed in \cite{shi18}, where $\Sigma_{\rm SFR}=10^{-4.76}(\Sigma_{\rm gas}\Sigma_{\rm stars}^{0.5})^{1.09}$ \citep[see also][]{shi11}. As a check on the KS relation for our sample, we evaluated the power law correlation for $\Sigma_{\rm SFR}(\Sigma_{\rm gas})$ for the strong and weak spiral case and for both together (not shown). Writing the coefficient and power again as $C$ and $D$, we get for the strong spirals: $C=-4.024\pm0.470$, $D=1.359\pm0.348$, for the weak spirals: $C=-4.019\pm0.564$, $D=1.295\pm0.510$, and for all together: $C=-4.067\pm0.424$, $D=1.373\pm0.337$. These three fits are all very similar to each other, illustrating the robustness of the KS relation, and the slopes are about the same as in the conventional KS relation for total gas, i.e., $\sim1.4$.
          
\begin{figure}
\includegraphics[width=10cm]{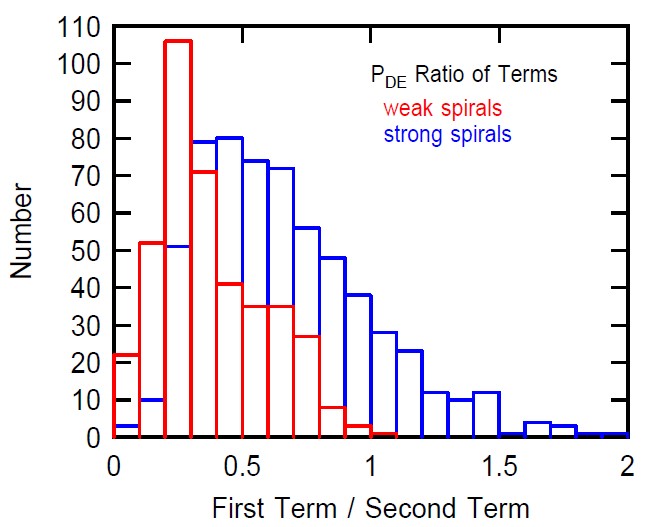}
\caption{Histograms of the ratio of the first term in $P_{\rm DE}$ to the second term for weak and strong arm spirals. }
\label{PDEhis}
\centering
\end{figure}

To investigate these two terms in $P_{\rm DE}$ further, we plot histograms in Figure \ref{PDEhis} of the ratio of the first to the second term for all the weak spiral (red) and strong spiral (blue) data. Generally the second term dominates \cite[see also][]{fisher19}, especially in galaxies with weak spiral arms. This means that the $\Sigma_{\rm SFR}(P_{\rm DE})$ relation is analogous to the Extended KS law, especially if $H_{\rm star}$ varies with radius more slowly than the surface densities. 

\begin{figure}
\includegraphics[width=12cm]{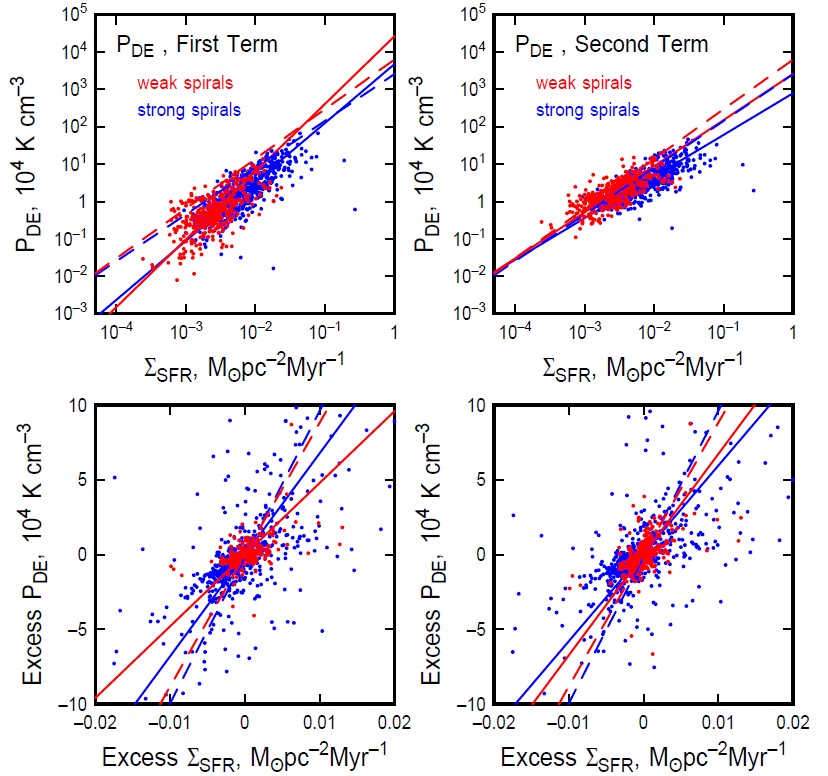}
\caption{(bottom) Correlations between azimuthal fluctuations in $\Sigma_{\rm SFR}$ and azimuthal fluctuations in each term of the dynamical pressure, with the first term on the left and the second term on the right. (top) Same for total parameter values. Fitting parameters are in Table \ref{table:PDEs}. }
\label{PDEs}
\centering
\end{figure}

Figure \ref{PDEs} plots the excess and total relationships between the individual components in $P_{\rm DE}$ and $\Sigma_{\rm SFR}$. The solid lines are bivariate fits for these components, and the dashed lines are fits to the relationships with the full $P_{\rm DE}$ values, from Figure \ref{pressure}. Fitted values are in Table \ref{table:PDEs}. The excess relationships for the individual terms (bottom panels) are each shallower than the excess relationship for the full $P_{\rm DE}$, for trivial reasons of arithmetic. That is, if $P_{\rm DE,1}=A_1+B_1\Sigma_{\rm SFR}$ and $P_{\rm DE,2}=A_2+B_2\Sigma_{\rm SFR}$ are the relationships for each term in $P_{\rm DE}$, then $P_{\rm DE}\equiv P_{\rm DE,1}+P_{\rm DE,2}=(A_1+A_2)+(B_1+B_2)\Sigma_{\rm SFR}$ is the relationship for the full $P_{\rm DE}$, having a slope equal to the sum of the slopes of the individual ones.  

We note from the top panels of Figure \ref{PDEs} that the first term in $P_{\rm DE}$ has a steeper power law relationship with $\Sigma_{\rm SFR}$ than the second term. In Table \ref{table:PDEs}, the powers for the first term are 1.58 and 1.81 for strong and weak spirals, respectively. If this were the KS relation, then $P_{\rm DE,1}\propto\Sigma_{\rm gas}^2$ would be proportional to $\Sigma_{\rm SFR}$ to the power 2/1.4=1.4. The slopes of the power-law relationships between the second term in $P_{\rm DE}$ and $\Sigma_{\rm SFR}$ are 1.11 and 1.23 for strong and weak spirals, which are shallower than the slopes of the total relationships in Figure \ref{pressure} and comparable to the slope of the Extended KS law when written with $\Sigma_{\rm SFR}$ as the dependent variable. 

The slopes of the excursions in the bottom panels of Figure \ref{PDEs}, written as $B$ in Table \ref{table:PDEs}, are consistent with the slopes derived from the derivatives of the power laws in the top panels, written as $B^\prime$ in Table \ref{table:PDEs}.  This implies that each term in $P_{\rm DE}$ always follows $\Sigma_{\rm SFR}$ closely, i.e., in radius and from galaxy to galaxy, and in azimuthal variations.  This is the same as the total $P_{\rm DE}$, where $B$ and $B^\prime$ were about the same in Table \ref{table:pressure}, but it differs from the molecular pressures, $P_{\rm O150}$ and $P_{\rm P150}$, which have larger excursions in azimuth than consistent with their global relationships (Sect. \ref{sect:pressure}).

The tight correlation between $\Sigma_{\rm SFR}$ and $P_{\rm DE}$, as well as with each term in $P_{\rm DE}$, seems consistent with feedback control of total pressure, but it has to do this without affecting molecular clouds much (Sect. \ref{fluctuations}), and the correlations are similar to the KS relation, which is not usually considered a result of feedback. If we think of $P_{\rm DE}$ as following $\Sigma_{\rm SFR}$, then feedback comes to mind, but if we think of the same relationship with $\Sigma_{\rm SFR}$ following the gas and star surface density in a KS relation, then cloud formation comes to mind. Is $P_{\rm DE}$, through its two terms that each resemble a KS relation, a precursor to star formation, or is $P_{\rm DE}$ a result of it because of feedback?

For an equilibrium model, both would be true, but star formation and gas surface density are only in equilibrium on sufficiently long spatial and temporal timescales. The close correlation for azimuthal variations found here, especially for strong spiral arms where they presumably reflect the arm/interarm contrast, suggest the components of $P_{\rm DE}$ are a precursor to $\Sigma_{\rm SFR}$.  For example, if most of the regions with locally high $\Sigma_{\rm SFR}$, which also have high molecular masses (Sect. \ref{fluctuations}), are too young for supernovae to have put in the full yield of feedback momentum, then the terms in $P_{\rm DE}$ would have to be precursors. This is consistent with the discussion in the Introduction about evidence for increased \HI\ velocity dispersions in regions of former star formation, 50-100 Myr ago \citep{stilp13,hunter22,hunter23}. If feedback takes that long to show up in the pressure, then the tight correlation found here for azimuthal fluctuations would seem to imply that star formation is following the gas mass, which is the $\Sigma_{\rm SFR}(P_{\rm DE,1})$ and $\Sigma_{\rm SFR}(P_{\rm DE,2})$ way of looking at it. 

Figure 3 in \cite{ostriker22} shows $P_{\rm DE}$ and $\Sigma_{\rm SFR}$ as functions of time. $\Sigma_{\rm SFR}$ fluctuates by factors of 10 to 100 over time in various models, while $P_{\rm DE}$ varies by less than a factor of 2. This is consistent with the small excursion in $P_{\rm DE}$ in the bottom-right panel of Figure \ref{pressure} and the bottom panels of Figure \ref{PDEs}. 

\section{Turbulence and Pressure from ISM Gravity and Spiral Dynamics}
\label{dynamics}

Aside from the question of whether the components of $P_{\rm DE}$ are a precursor or a result of star formation, a different consideration is that a third process, such as spiral arm dynamics or shear, might drive them both. These processes act on the total gas at a rate that depends on local galaxy dynamics. 

For example, if the effective Toomre parameter $Q$ is regulated by galactic dynamics to be about constant, which seems to be the case \citep{romeo17}, then the velocity dispersion is partially regulated by swing amplification and local gravitational collapse. The effective dispersion, considering also magnetic and cosmic ray pressures that contribute to the stability of the ISM, is then given by
$\sigma_{\rm Q}={{\pi G \Sigma_{\rm Q} Q}/{\kappa}}$ for some effective surface density $\Sigma_{\rm Q}$ and effective $Q$, generally involving both stars and gas,  and for epicyclic frequency $\kappa$. Combining this with the disk thickness, $H=\sigma_{\rm Q}^2/(\pi G \Sigma_{\rm H})$, where $\Sigma_{\rm H}$ is the combined gas, star, and dark matter surface density inside the gas layer, and converting the resultant midplane gas density, $\rho_{\rm gas}=0.5\Sigma_{\rm gas}/H$, to a dynamical rate $(G\rho_{\rm gas})^{0.5}$ for total gas surface density $\Sigma_{\rm gas}$, gives the expectation that
\begin{equation}
\Sigma_{\rm SFR}\sim\epsilon_\rho\Sigma_{\rm gas}(G\rho_{\rm gas})^{0.5}=\epsilon_{\rm \rho}\Sigma_{\rm gas}\kappa\left({{\Sigma_{\rm gas}\Sigma_{\rm H}}\over{2\pi\Sigma_{\rm Q}^2Q^2}}\right)^{0.5}\sim\epsilon_\kappa\Sigma_{\rm gas}\kappa .
\label{eq:sfr}
\end{equation}
The square root term was absorbed into the generic efficiency for this density-based expression, $\epsilon_{\rm \rho}$, as a global efficiency, $\epsilon_\kappa$. In general, we would expect also
\begin{equation}
\Sigma_{\rm SFR}\sim\epsilon_{\rm A} \Sigma_{\rm gas} A \sim \epsilon_\Omega\Sigma_{\rm gas}\Omega
\label{eq:sfr2}
\end{equation}
for various other dynamical rates, the Oort rate of shear, $A$, and the orbital rate, $\Omega$. Conversions between these gas dynamical processes and the SFR involve the dimensionless efficiencies, $\epsilon$. 

\begin{figure}
\includegraphics[width=16cm]{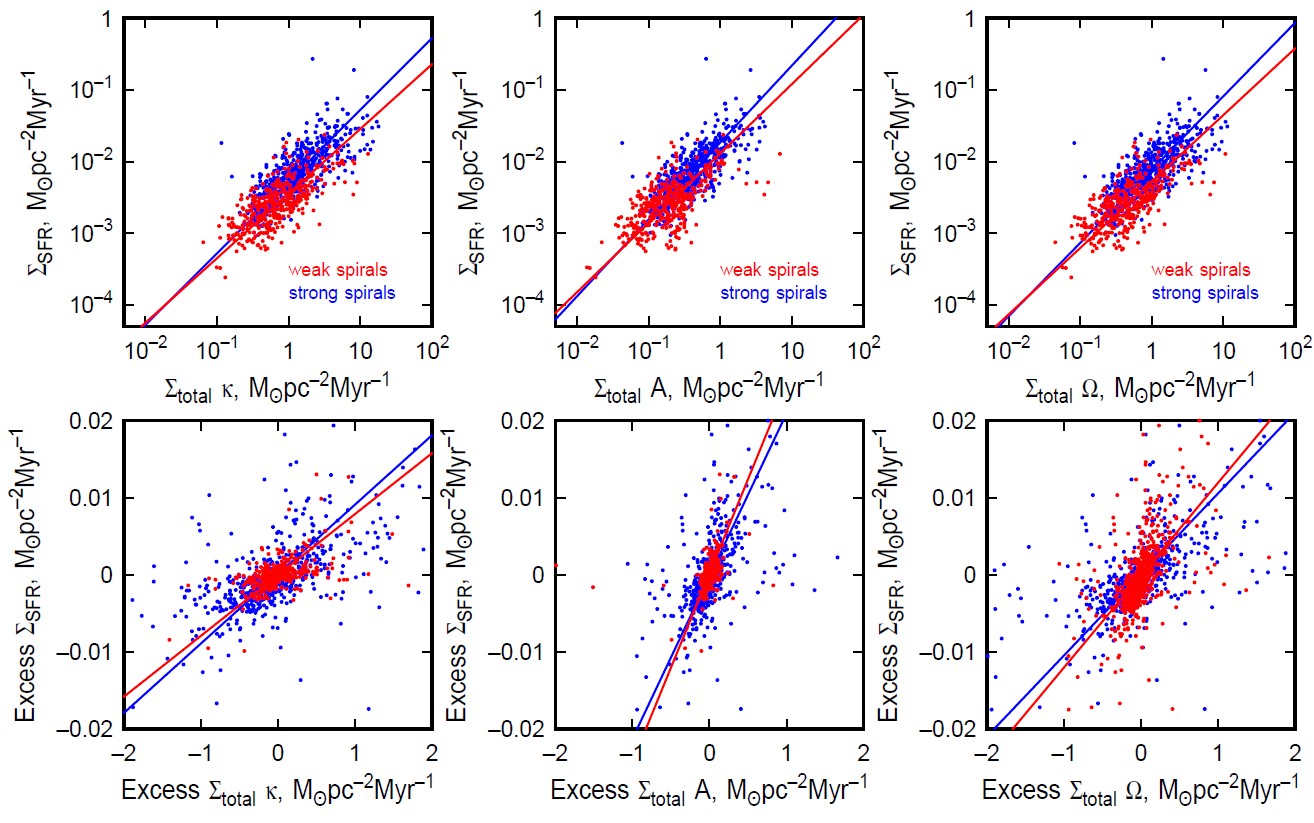}
\caption{Correlations between azimuthal fluctuations in the bottom row and total values in the top row for $\Sigma_{\rm SFR}$ as a function of various gas dynamical rates on a galactic scale, namely the epicycle rate $\kappa$, shear rate $A$, and orbital rate $\Omega$. The red and blue points and fitting lines are for galaxies with weak and strong spiral arms. Fitting parameters are in Table \ref{table:SFR}. }
\label{fig:SFR}
\centering
\end{figure}

Figure \ref{fig:SFR} plots the quantities given by equations \ref{eq:sfr} and \ref{eq:sfr2} for both total values (top panels) and excursions around the azimuth (bottom panels). We use the PHANGS tabulations for our 34 galaxies with $\Sigma_{\rm gas}=\Sigma_{\rm mol}+\Sigma_{\rm atom}$, again divided into 17 strong-arm and 17 weak-arm cases. The dynamical rates come from the PHANGS circular velocities, galactocentric radii, and rotation curve slopes, $\beta$.  The total correlations are fitted to power laws and the excursion relations fitted linearly, with fitting parameters in Table \ref{table:SFR}.  The slope of the excursion relation, $B$, has been multiplied by $100$ in the Table for clarity (equivalent to making the efficiency in units of 0.01 for these excursions). The coefficient of the power law fit, $C$, has not assumed this efficiency normalization; to do so would add 2 to each $C$.   

The slopes of the power law correlations, $D$, in Table \ref{table:SFR} are essentially 1.0, meaning that $\Sigma_{\rm SFR}$ tracks well the total gas surface density multiplied by a local galactic dynamical rate. All the fits are about the same, so there is no preferred rate indicated. This unity slope should not be compared with analogous correlations in \cite{sun23}, who got a sublinear slope when plotting $\Sigma_{\rm SFR}$ versus $\Sigma_{\rm mol}\Omega$. The main difference is that we use the total gas because that is what galactic-scale dynamical processes act upon. The efficiency factors, $\epsilon$, are given by $10^C$ for power law coefficients $C$ in Table \ref{table:SFR}. They are 0.52\%, 1.84\% and 0.78\% for dynamical rates $\kappa$, $A$ and $\Omega$. 

The slopes of the excursions in Table \ref{table:SFR}, listed as $B$, average to a factor of 1.8 more than the  derivatives of the power-law relations, listed as $B^\prime$. This factor is about that same as for $P_{\rm O150}$ and $P_{\rm P150}$ in Table \ref{table:pressure}, and may be for the same reason: star formation scales with gas density more sensitively in azimuthal perturbations (e.g., spiral arms) than the overall average. This implies a local factor of $\sim2$ enhancement of the star formation efficiency per unit total gas in spiral arms.

\section{Conclusions}
\label{conclusions}

Correlations between the surface density of the star formation rate, $\Sigma_{\rm SFR}$, and various properties of molecular and total gas were evaluated for 17 galaxies with strong spiral arms and 17 galaxies with weak arms. Arm strength was determined by the amplitude of the excursions in $\Sigma_{\rm SFR}$ measured around in azimuth. Data from the PHANGS survey were used. 

Azimuthal variations in $\Sigma_{\rm SFR}$ closely follow the azimuthal variations in molecular cloud mass, suggesting a fairly uniform star formation efficiency on the scale of 1.5 kpc. The CO velocity dispersion also follows the molecular cloud mass according to the usual cloud scaling laws. Thus we observe an increase in CO velocity dispersion with increased star formation rate, but this is the result of the usual scaling laws between velocity dispersion and molecular cloud mass prior to star formation, rather than the result of feedback from star formation affecting the molecular velocity dispersion (Sects. \ref{fluctuations} and \ref{total}).

Azimuthal variations in $\Sigma_{\rm SFR}$ are larger for given internal cloud pressure variations in strong arm galaxies than weak arm galaxies, suggesting that molecular clouds of the same pressure produce a higher star formation rate in strong spiral arms than in weak arms. An alternate interpretation is that molecular cloud pressure (e.g., from feedback) is higher for a given excursion in star formation rate in a weak arm galaxy than a strong arm galaxy (Sect. \ref{sect:pressure}).

Azimuthal variations in internal molecular cloud pressure are several times higher in proportion to star formation rate variations than are the total cloud pressure variations with the total star formation rate. This could be a result of background stars generally influencing the total cloud pressure over a wide range of galactic radii and galaxy types, while cloud pressure variations at a given radius for a particular galaxy are more affected by internal cloud gravity. In contrast, azimuthal variations in the total dynamical equilibrium pressure, $P_{\rm DE}$, follow variations in $\Sigma_{\rm SFR}$ in the same way as the total $P_{\rm DE}$ follows the total $\Sigma_{\rm SFR}$, suggesting a tight relation between the two quantities in all environments (Sect. \ref{sect:pressure}).

The internal molecular cloud pressure varies in azimuth ten times more strongly with the star formation rate than the dynamical equilibrium pressure does, and the internal cloud pressure is ten times higher than this equilibrium pressure for high star formation rates, reflecting the self-gravitational contribution to internal cloud pressure (Sect. \ref{sect:pressure}).

The dynamical equilibrium pressure is composed of two terms, with a first depending only on the total gas surface density and a second depending on this surface density as well as the volume stellar density. The first term alone would be analogous to the usual Kennicutt-Schmidt relation, although with a steeper power law since the square of $\Sigma_{\rm gas}$ enters rather than the 1.4 power in KS. The second term alone is analogous to the Extended KS relation, especially if the galactic disk thickness is assumed to vary less than the surface density. Thus, the good correlation between $P_{\rm DE}$ and $\Sigma_{\rm SFR}$ could be viewed either as a result of star formation feedback pressurizing the interstellar gas to an equilibrium value, or star formation following the presence of interstellar gas, with a collapse rate supplemented by the background stellar density. The physical origin of the relationship is interpreted differently if we write $\Sigma_{\rm SFR}$ or $P_{\rm DE}$ as the dependent variable. The tightness of the correlation between them, i.e., for all radial and galactic variations and for all azimuthal variations, combined with the observation and concept that feedback acts on the gas over a significant time span, such as 50 to 100 Myr, while collapse to star formation seems to follow quickly and universally from the presence of dense gas, suggests an interpretation in which excess gas leads rapidly to excess star formation, and $P_{\rm DE}$ is primarily a proxy for the presence of gas. 

The star formation rate scales linearly with the product of the total gas surface density and various galactic rates, i.e., the epicyclic rate, the shear rate, and the orbital rate, suggesting an important role for galactic dynamics in controlling star formation (Sect. \ref{dynamics}).

Helpful comments by the referee are acknowledged.

\newpage
\begin{deluxetable}{ccccc}
\tabletypesize{\scriptsize}
\tablecaption{Galaxy Sample, $M_{\rm star}>2\times10^{10}\;M_\odot$  \label{table:sample}}
\tablewidth{0pt}
\tablehead{
\multicolumn{2}{c}{Broad $\Delta\Sigma_{\rm SFR}$\tablenotemark{a}}&
\multicolumn{2}{c}{Narrow $\Delta\Sigma_{\rm SFR}$\tablenotemark{b}}&
\colhead{Too Few CO Obs.}
}
\startdata
NGC 628 & NGC 1559 & NGC  253 & NGC 1097 & ESO 97-G13\\
NGC 1566 & NGC 1792 & NGC 1300 & NGC 1365 & NGC 1317\\
NGC 2566 & NGC 2903 & NGC 1433 & NGC 1512 & NGC 3626\\
NGC 2997 & NGC 3059 & NGC 1546 & NGC 2775 & NGC 4293\\
NGC 3627 & NGC 4254 & NGC 3351 & NGC 3507 & NGC 4459\\
NGC 4303 & NGC 4321 & NGC 3521 & NGC 4457 & NGC 4477\\
NGC 4535 & NGC 5236 & NGC 4536 & NGC 4548 & NGC 4596\\
NGC 5248 & NGC 5643 & NGC 4569 & NGC 5134 & NGC 5128\\
NGC 6300 &          & NGC 6744 &          & NGC 7743\\
\enddata
\tablenotetext{\textrm{a}}{Broad $\Sigma_{\rm SFR}$ typically corresponds to galaxies with strong spiral arms}
\tablenotetext{\textrm{b}}{Narrow $\Sigma_{\rm SFR}$ typically corresponds to galaxies with weak spiral arms or flocculent arms}
\end{deluxetable}

\begin{deluxetable}{lcccc}
\tabletypesize{\scriptsize}
\tablecaption{Linear Fits\tablenotemark{a} to Excursion Correlations, $\Delta y=A+B\Delta x$, in Figures \ref{all-pix}, \ref{all-obj}, \ref{broad-full-mass} and \ref{narrow-full-mass}  \label{table:linearfits2}}
\tablewidth{0pt}
\tablehead{
\colhead{Galaxy Type/}
&\multicolumn{2}{c}{Pixels}
&\multicolumn{2}{c}{Objects}\\
\colhead{Correlation}
&\colhead{A}&\colhead{B}
&\colhead{A}&\colhead{B}
}
\startdata
{\bf Strong Spirals}&&&&\\
$\Delta\sigma(\Delta\Sigma_{\rm SFR})$&
$     0.079\pm     0.082$&$   324\pm   188$&
$     0.125\pm     0.114$&$   349\pm   259$\\
$\Delta\sigma(\Delta\Sigma_{\rm mol}[\Delta\Sigma_{\rm SFR}])$\tablenotemark{b}
&0.084&310&0.124&338\\
$\Delta\sigma(\Delta M_{\rm mol}[\Delta\Sigma_{\rm SFR}])$\tablenotemark{c}
&&&0.103&323\\
$\Delta\sigma(\Delta\Sigma_{\rm mol})$
&$     0.006\pm     0.009$&$     0.047\pm     0.008$&
$     0.112\pm     0.092$&$     0.027\pm     0.017$\\
$\Delta\Sigma_{\rm mol}(\Delta\Sigma_{\rm SFR})$
&$     1.65\pm     1.23$&$  6600\pm  3090$&
$     0.451\pm     3.377$&$ 12800\pm  6480$\\
$\Delta\alpha_{\rm vir}(\Delta\Sigma_{\rm SFR})$
&$    -0.098\pm     0.084$&$  -240\pm   217$&
$    -0.150\pm     0.104$&$  -287\pm   254$\\
$\Delta\sigma(\Delta M_{\rm mol})$
&&&$0.164\pm     0.126$&$     3.39\pm     1.62$\\
$\Delta M_{\rm mol}(\Delta\Sigma_{\rm SFR})$
&&&$    -0.018\pm     0.036$&$    95.5\pm    57.5$\\

{\bf Weak Spirals}&&&&\\

$\Delta\sigma(\Delta\Sigma_{\rm SFR})$&
$    -0.062\pm     0.006$&$   871\pm   680$&
$    -0.034\pm     0.002$&$  1220\pm  1120$\\
$\Delta\sigma(\Delta\Sigma_{\rm mol}[\Delta\Sigma_{\rm SFR}])$\tablenotemark{b}
&-0.030&876&-0.033&2270\\
$\Delta\sigma(\Delta M_{\rm mol}[\Delta\Sigma_{\rm SFR}])$\tablenotemark{c}
&&&0.036&2650\\
$\Delta\sigma(\Delta\Sigma_{\rm mol})$
&$     0.004\pm     0.018$&$     0.086\pm     0.030$&
$     0.021\pm     0.042$&$     0.068\pm     0.055$\\
$\Delta\Sigma_{\rm mol}(\Delta\Sigma_{\rm SFR})$
&$    -0.394\pm     0.105$&$ 10200\pm  6830$&
$    -0.782\pm     0.131$&$ 33200\pm 31000$\\
$\Delta\alpha_{\rm vir}(\Delta\Sigma_{\rm SFR})$
&$     0.009\pm     0.013$&$ -9300\pm  9290$&
$    -0.052\pm     0.020$&$ -1890\pm  1850$\\
$\Delta\sigma(\Delta M_{\rm mol})$
&&&$    -0.007\pm     0.022$&$     6.31\pm     4.08$\\
$\Delta M_{\rm mol}(\Delta\Sigma_{\rm SFR})$
&&&$     0.007\pm     0.010$&$   419\pm   404$\\
\enddata
\tablenotetext{\textrm{a}}{Numbers rounded to 3 significant figures in all tables. 
}
\tablenotetext{\textrm{b}}{Consistency check in the proposed secondary correlation, $\Delta\sigma(\Delta\Sigma_{\rm SFR})$, using primary correlations
$\Delta\sigma(\Delta\Sigma_{\rm mol})$ and $\Delta\Sigma_{\rm mol}(\Delta\Sigma_{\rm SFR})$, which occur for both pixel and object measurements (Figs. 3,4).
}
\tablenotetext{\textrm{c}}{Consistency check in the proposed secondary correlation, $\Delta\sigma(\Delta\Sigma_{\rm SFR})$, using primary correlations
$\Delta\sigma(\Delta M_{\rm mol})$ and $\Delta M_{\rm mol}(\Delta\Sigma_{\rm SFR})$, which exist only for object measurements (Figs. 8,9).
}
\end{deluxetable}

\begin{deluxetable}{lcccc}
\tabletypesize{\scriptsize}
\tablecaption{Power Law Fits $y=Cx^D$ to Total Correlations in Figures \ref{all-full-pix}, \ref{all-full-obj}, \ref{broad-full-mass} and \ref{narrow-full-mass} \label{table:powerlawfits}}
\tablewidth{0pt}
\tablehead{
\colhead{Galaxy Type/}
&\multicolumn{2}{c}{Pixels}
&\multicolumn{2}{c}{Objects}\\
\colhead{Correlation}
&\colhead{C}&\colhead{D}
&\colhead{C}&\colhead{D}
}
\startdata
{\bf Strong Spirals}&&&&\\
$\sigma(\Sigma_{\rm SFR})$
&$     1.66\pm     0.28$&$     0.471\pm     0.127$
&$     1.59\pm     0.42$&$     0.394\pm     0.194$\\
$\sigma(\Sigma_{\rm mol})$
&$     0.051\pm     0.055$&$     0.425\pm     0.041$
&$     0.022\pm     0.358$&$     0.410\pm     0.207$\\
$\Sigma_{\rm mol}(\Sigma_{\rm SFR})$
&$     3.79\pm     0.57$&$     1.103\pm     0.254$
&$     3.99\pm     0.76$&$     1.03\pm     0.35$\\
$\alpha_{\rm vir}(\Sigma_{\rm SFR})$\tablenotemark{a}
&$    14.1\pm    11.2$&$     5.45\pm     5.02$
&$   -25.2\pm    26.1$&$   -12.2\pm    11.9$\\
$\sigma(M_{\rm mol})$
&&&$    -1.60\pm     0.73$&$     0.351\pm     0.110$\\
$M_{\rm mol}(\Sigma_{\rm SFR})$
&&&$     9.117\pm     0.964$&$     1.13\pm     0.44$\\
{\bf Weak Spirals}&&&&\\
$\sigma(\Sigma_{\rm SFR})$
&$     2.23\pm     0.70$&$     0.646\pm     0.266$
&$     2.08\pm     0.92$&$     0.546\pm     0.359$\\
$\sigma(\Sigma_{\rm mol})$
&$     0.045\pm     0.062$&$     0.495\pm     0.065$
&$    -0.030\pm     0.415$&$     0.496\pm     0.292$\\
$\Sigma_{\rm mol}(\Sigma_{\rm SFR})$
&$     4.28\pm     0.96$&$     1.25\pm     0.37$
&$     4.42\pm     1.27$&$     1.17\pm     0.50$\\
$\alpha_{\rm vir}(\Sigma_{\rm SFR})$\tablenotemark{a}
&$    22.1\pm    19.1$&$     7.66\pm     7.25$
&$  -659\pm   660$&$  -258\pm   258$\\
$\sigma(M_{\rm mol})$
&&&$    -2.06\pm     1.36$&$     0.426\pm     0.212$\\
$M_{\rm mol}(\Sigma_{\rm SFR})$
&&&$     9.99\pm     1.94$&$     1.39\pm     0.75$\\
\enddata
\tablenotetext{\textrm{a}}{This was fit to a linear-log relation, $\alpha_{\rm vir}=C+D\log(\Sigma_{\rm SFR})$, as shown in the figures.}
\end{deluxetable}

\begin{deluxetable}{lccccc}
\tabletypesize{\scriptsize}
\tablecaption{Comparison Between the Excess Correlation Slope and Predictions from the Power Law Fits to the Total Correlations \label{table:comparisons}}
\tablewidth{0pt}
\tablehead{
\colhead{Data Type}
&\colhead{Excess Correlation}
&\colhead{Power Law }
&\multicolumn{2}{c}{Average Values}
&\colhead{Predicted Excess Correlation}\\
\colhead{}
&\colhead{Slope\tablenotemark{a}, $B_{\sigma,{\rm SFR}}$}
&\colhead{Slope\tablenotemark{b}, $D_{\sigma,{\rm SFR}}$}
&\colhead{$<\sigma>$}
&\colhead{$<\Sigma_{\rm SFR}>$}
&\colhead{Slope, $B_{\sigma,{\rm SFR}}^\prime$}\\
&&&km s$^{-1}$
&$M_\odot$ pc$^{-2}$ Myr$^{-1}$
&
}
\startdata
Strong Spirals, Pixels&324&0.471&4.11&$5.85\times10^{-3}$&330\\
Strong Spirals, Objects&349&0.394&5.41&$6.56\times10^{-3}$&324\\
Weak Spirals, Pixels&871&0.646&3.43&$2.37\times10^{-3}$&936\\
Weak Spirals, Objects&1220&0.546&4.81&$2.73\times10^{-3}$&960
\enddata
\tablenotetext{\textrm{a}}{From Table \ref{table:linearfits2}.}
\tablenotetext{\textrm{b}}{From Table \ref{table:powerlawfits}.}
\end{deluxetable}

\begin{deluxetable}{lcccc}
\tabletypesize{\scriptsize}
\tablecaption{Linear Fits to Excursion Correlations ($\Delta y=A+B\Delta x$) and Power-Law Fits to Total Correlations ($\log y=C+D\log x$)
between $P$\tablenotemark{a} and $\Sigma_{\rm SFR}$\tablenotemark{b} in Figure \ref{pressure} \label{table:pressure}}
\tablewidth{0pt}
\tablehead{
\colhead{Galaxy Type/}
&\colhead{A\tablenotemark{c}}&\colhead{B\tablenotemark{c}}
&\colhead{C\tablenotemark{d}}&\colhead{D\tablenotemark{d}}\\
\colhead{Correlation}&\colhead{}&\colhead{}&\colhead{}&\colhead{}
}
\startdata
{\bf Strong Spirals, objects}&&&&\\
$\Delta P_{\rm O150}(\Delta\Sigma_{\rm SFR})$
&$    -0.73\pm       2.68$&$  7210\pm    4530$&&\\

$B_{\rm O150}^\prime (C_{\rm O150},D_{\rm O150})$&&3330&&\\

$P_{\rm O150}(\Sigma_{\rm SFR})$
&&&$     4.61\pm       1.11$&$     1.59\pm       0.51$\\

{\bf Strong Spirals, pixels}&&&&\\

$\Delta P_{\rm P150}(\Delta\Sigma_{\rm SFR})$
&$    -0.41\pm       1.75$&$  5730\pm    3560$&&\\

$B_{\rm P150}^\prime (C_{\rm P150},D_{\rm P150})$&&1830&&\\

$P_{\rm P150}(\Sigma_{\rm SFR})$
&&&$      5.20\pm       1.04$&$     2.00\pm       0.47$\\

$\Delta P_{\rm DE}(\Delta\Sigma_{\rm SFR})$
&$    -0.154\pm       0.220$&$   979\pm     544$&&\\

$B_{\rm DE}^\prime (C_{\rm DE},D_{\rm DE})$&&909&&\\

$P_{\rm DE}(\Sigma_{\rm SFR})$
&&&$     3.41\pm       0.66$&$     1.25\pm       0.30$\\

{\bf Weak Spirals, objects}&&&&\\

$\Delta P_{\rm O150}(\Delta\Sigma_{\rm SFR})$
&$    -2.53\pm       0.56$&$ 16900\pm   15600$&&\\

$B_{\rm O150}^\prime (C_{\rm O150},D_{\rm O150})$&&4180&&\\

$P_{\rm O150}(\Sigma_{\rm SFR})$
&&&$     5.69\pm       2.24$&$     1.92\pm       0.88$\\

{\bf Weak Spirals, pixels}&&&&\\

$\Delta P_{\rm P150}(\Delta\Sigma_{\rm SFR})$
&$    -1.15\pm       0.21$&$ 15500\pm   12800$&&\\

$B_{\rm P150}^\prime (C_{\rm P150},D_{\rm P150})$&&2090&&\\

$P_{\rm P150}(\Sigma_{\rm SFR})$
&&&$    6.85\pm       2.21$&$     2.50\pm       0.84$\\

$\Delta P_{\rm DE}(\Delta\Sigma_{\rm SFR})$
&$-0.000\pm       0.009$&$   882\pm     475$&&\\

$B_{\rm DE}^\prime (C_{\rm DE},D_{\rm DE})$&&1180&&\\

$P_{\rm DE}(\Sigma_{\rm SFR})$
&&&$     3.80\pm       1.05$&$     1.33\pm       0.40$\\

\enddata
\tablenotetext{\textrm{a}}{Pressure $P$ in units of $10^4$ K cm$^{-3}$}
\tablenotetext{\textrm{b}}{Star formation rate density, $\Sigma_{\rm SFR}$, in units of $M_\odot$ pc$^{-2}$ Myr$^{-1}$}
\tablenotetext{\textrm{c}}{$A$ and $B$ are such that $\Delta P=A+B\Delta\Sigma_{\rm SFR}$.}
\tablenotetext{\textrm{d}}{$C$ and $D$ are such that $\log P=C+D\log \Sigma_{\rm SFR}$, or $P=10^C \Sigma_{\rm SFR}^D$.}
\end{deluxetable}

\begin{deluxetable}{lcccc}
\tabletypesize{\scriptsize}
\tablecaption{Linear Fits to Excursion Correlations ($\Delta y=A+B\Delta x$) and Power-Law Fits to Total Correlations ($\log y=C+D\log x$)
between 1st and 2nd terms of $P_{\rm DE}$\tablenotemark{a} and $\Sigma_{\rm SFR}$\tablenotemark{b} in Figure \ref{PDEs} \label{table:PDEs}}
\tablewidth{0pt}
\tablehead{
\colhead{Galaxy Type/}
&\colhead{A\tablenotemark{c}}&\colhead{B\tablenotemark{c}}
&\colhead{C\tablenotemark{d}}&\colhead{D\tablenotemark{d}}\\
\colhead{Correlation}&\colhead{}&\colhead{}&\colhead{}&\colhead{}
}
\startdata
{\bf Strong Spirals, pixels}&&&&\\
$\Delta P_{\rm DE,1}(\Delta\Sigma_{\rm SFR})$
&$    -0.015\pm       0.142$&$   678\pm     339$&&\\

$B_{\rm DE,1}^\prime (C_{\rm DE,1},D_{\rm DE,1})$&&410&&\\

$P_{\rm DE,1}(\Sigma_{\rm SFR})$
&&&$     3.68\pm       0.89$&$     1.58\pm       0.40$\\

$\Delta P_{\rm DE,2}(\Delta\Sigma_{\rm SFR})$
&$     0.0775\pm       0.0882$&$   587\pm     297$&&\\

$B_{\rm DE,2}^\prime (C_{\rm DE,2},D_{\rm DE,2})$&&496&&\\

$P_{\rm DE,2}(\Sigma_{\rm SFR})$
&&&$     2.90\pm       0.61$&$     1.11\pm       0.28$\\

{\bf Weak Spirals, pixels}&&&&\\

$\Delta P_{\rm DE,1}(\Delta\Sigma_{\rm SFR})$
&$    -0.00001\pm       0.00000$&$   480\pm     296$&&\\

$B_{\rm DE,1}^\prime (C_{\rm DE,1},D_{\rm DE,1})$&&384&&\\

$P_{\rm DE,1}(\Sigma_{\rm SFR})$
&&&$     4.42\pm       1.86$&$     1.81\pm       0.72$\\

$\Delta P_{\rm DE,2}(\Delta\Sigma_{\rm SFR})$
&$    -0.00000\pm       0.00000$&$   673\pm     400$&&\\

$B_{\rm DE,2}^\prime (C_{\rm DE,2},D_{\rm DE,2})$&&798&&\\

$P_{\rm DE,2}(\Sigma_{\rm SFR})$
&&&$     3.42\pm       0.93$&$     1.23\pm       0.36$\\

\enddata
\tablenotetext{\textrm{a}}{Pressure $P$ in units of $10^4$ K cm$^{-3}$}
\tablenotetext{\textrm{b}}{Star formation rate density, $\Sigma_{\rm SFR}$, in units of $M_\odot$ pc$^{-2}$ Myr$^{-1}$}
\tablenotetext{\textrm{c}}{$A$ and $B$ are such that $\Delta P=A+B\Delta\Sigma_{\rm SFR}$.}
\tablenotetext{\textrm{d}}{$C$ and $D$ are such that $\log P=C+D\log \Sigma_{\rm SFR}$, or $P=10^C \Sigma_{\rm SFR}^D$.}
\end{deluxetable}

\begin{deluxetable}{lcccc}
\tabletypesize{\scriptsize}
\tablecaption{Linear Fits to Excursion Correlations ($\Delta y=A+B\Delta x$) and Power-Law Fits to Total Correlations ($\log y=C+D\log x$) between $\Sigma_{\rm SFR}$\tablenotemark{a} and various surface density rates in Figure \ref{fig:SFR}  \label{table:SFR}}
\tablewidth{0pt}
\tablehead{
\colhead{Galaxy Type/}
&\colhead{A\tablenotemark{b}}&\colhead{B\tablenotemark{b}}
&\colhead{C\tablenotemark{c}}&\colhead{D\tablenotemark{c}}\\
\colhead{Correlation}&\colhead{}&\colhead{$(\times100$)}&\colhead{}&\colhead{}
}
\startdata
{\bf Strong Spirals, pixels}&&&&\\

$\Delta \Sigma_{\rm SFR}(\Delta\Sigma_{\rm gas}\kappa)$
&$     0.00008\pm       0.00024$&$     0.902\pm       0.474$&&\\

$B_{\rm SFR}^\prime (C_{S,\kappa},D_{S,\kappa})$&&0.525&&\\

$\Delta \Sigma_{\rm SFR}(\Delta\Sigma_{\rm gas}A)$
&$    -0.00019\pm       0.00005$&$     2.11\pm       1.21$&&\\

$B_{\rm SFR}^\prime (C_{SA},D_{SA})$&&1.84&&\\

$\Delta \Sigma_{\rm SFR}(\Delta\Sigma_{\rm gas}\Omega)$
&$     0.00004\pm       0.00018$&$     1.05\pm       0.53$&&\\

$B_{\rm SFR}^\prime (C_{S\Omega},D_{S,\Omega})$&&0.79&&\\

$\Sigma_{\rm SFR}(\Sigma_{\rm gas}\kappa)$
&&&$    -2.28\pm       0.02$&$     1.00\pm       0.26$\\

$\Sigma_{\rm SFR}(\Sigma_{\rm gas}A)$
&&&$    -1.73\pm       0.11$&$     1.07\pm       0.26$\\

$\Sigma_{\rm SFR}(\Sigma_{\rm gas}\Omega)$
&&&$    -2.11\pm       0.02$&$     1.02\pm       0.26$\\

{\bf Weak Spirals, pixels}&&&&\\

$\Delta \Sigma_{\rm SFR}(\Delta\Sigma_{\rm gas}\kappa)$
&$    -0.00004\pm       0.00001$&$     0.79\pm       0.46$&&\\

$B_{\rm SFR}^\prime (C_{S,\kappa},D_{S,\kappa})$&&0.334&&\\

$\Delta \Sigma_{\rm SFR}(\Delta\Sigma_{\rm gas}A)$
&$     0.00017\pm       0.00015$&$     2.45\pm       2.09$&&\\

$B_{\rm SFR}^\prime (C_{SA},D_{SA})$&&1.32&&\\

$\Delta \Sigma_{\rm SFR}(\Delta\Sigma_{\rm gas}\Omega)$
&$    -0.00004\pm       0.00002$&$     1.20\pm       0.70$&&\\

$B_{\rm SFR}^\prime (C_{S\Omega},D_{S,\Omega})$&&0.52&&\\

$\Sigma_{\rm SFR}(\Sigma_{\rm gas}\kappa)$
&&&$    -2.45\pm       0.05$&$     0.903\pm       0.311$\\

$\Sigma_{\rm SFR}(\Sigma_{\rm gas}A)$
&&&$    -1.88\pm       0.31$&$     0.968\pm       0.429$\\

$\Sigma_{\rm SFR}(\Sigma_{\rm gas}\Omega)$
&&&$    -2.28\pm       0.11$&$     0.928\pm       0.335$\\

\enddata
\tablenotetext{\textrm{a}}{Star formation rate density, $\Sigma_{\rm SFR}$, and surface density rates are in units of $M_\odot$ pc$^{-2}$ Myr$^{-1}$}
\tablenotetext{\textrm{b}}{$A$ and $B$ are such that $\Delta \Sigma_{\rm SFR}=A+B(\Delta\Sigma_{\rm gas}R)$ for rate $R$. $B$ values multiplied by 100.}
\tablenotetext{\textrm{c}}{$C$ and $D$ are such that $\log \Sigma_{\rm SFR}=C+D\log (\Sigma_{\rm gas}R)$, or $\Sigma_{\rm SFR}=10^C (\Sigma_{\rm gas}R)^D$ for rate $R$.}
\end{deluxetable}

\end{document}